 \documentclass[a4paper,11pt]{article}
\usepackage{jheppub}

\usepackage{placeins}
\usepackage{mathrsfs}
\usepackage{wasysym}
\usepackage{verbatim}
\usepackage{booktabs}
\usepackage[tight]{subfigure}
\usepackage{color}
\usepackage{graphicx}
\usepackage{siunitx}

\def\TeV{\, {\rm TeV}}
\def\GeV{\, {\rm GeV}}

\def\ie{\emph{i.e.}}
\newcommand{\be}{\begin{equation}}
\newcommand{\ee}{\end{equation}}
\newcommand{\bea}{\begin{eqnarray}}
\newcommand{\eea}{\end{eqnarray}}

\newcommand{\df}{\dfrac}

\def\simlt{\stackrel{<}{{}_\sim}}
\def\simgt{\stackrel{>}{{}_\sim}}
\newcommand{\gsim}{\lower.7ex\hbox{$\;\stackrel{\textstyle>}{\sim}\;$}}
\newcommand{\lsim}{\lower.7ex\hbox{$\;\stackrel{\textstyle<}{\sim}\;$}}

\def\mstl{m_{\tilde t_{1}}}
\def\msth{m_{\tilde t_{2}}}

\def\mass2{mass${}^2$}

\def\mass2{mass${}^2$}

\def\simlt{\stackrel{<}{{}_\sim}}
\def\simgt{\stackrel{>}{{}_\sim}}

\definecolor{nicered}{rgb}{0.7,0.1,0.1}
\definecolor{nicegreen}{rgb}{0.1,0.5,0.1}
\hypersetup{colorlinks,citecolor= nicegreen,linkcolor= nicered}

\title{ What is a Natural SUSY scenario?}
\author[a]{J. Alberto Casas,} \author[a]{Jes\'us M. Moreno,} \author[a,b]{Sandra Robles,} \author[a]{Krzysztof Rolbiecki} \author[c]{and Bryan Zald\'ivar}
\affiliation[a]{Instituto de F\'isica Te\'orica, IFT-UAM/CSIC, Universidad Aut\'onoma de Madrid, \\
        Cantoblanco, 28049 Madrid, Spain }
\affiliation[b]{Departamento de F\'{\i}sica Te\'{o}rica, Universidad Aut\'{o}noma de Madrid,  \\
        Cantoblanco, 28049 Madrid, Spain }
\affiliation[c]{Service de Physique Th\'eorique, Universit\'e Libre de Bruxelles, \\
         Boulevard du Triomphe, CP225, 1050 Brussels, Belgium
         }
\emailAdd{alberto.casas@uam.es} \emailAdd{jesus.moreno@csic.es} \emailAdd{sandra.robles@uam.es} \emailAdd{krzysztof.rolbiecki@desy.de} \emailAdd{bryan.zaldivar@ulb.ac.be} 

\abstract{\small

The idea of ``Natural SUSY", understood as a supersymmetric scenario where the fine-tuning is as mild as possible, is a reasonable guide to explore supersymmetric phenomenology.
In this paper, we re-examine this issue in the context of the MSSM including several improvements, such as the mixing of the fine-tuning conditions for different soft terms and the presence of potential extra fine-tunings that must be combined with the electroweak one. We give tables and plots that allow to easily evaluate the fine-tuning and the corresponding naturalness bounds for any theoretical model defined at any high-energy (HE) scale. Then, we analyze in detail the complete fine-tuning bounds for the unconstrained MSSM, defined at any HE scale. We show that Natural SUSY does {\em not} demand light stops. Actually, an average stop mass below 800~GeV is disfavored, though one of the stops might be very light. 
Regarding phenomenology, the most stringent upper bound from naturalness is the one on the gluino mass, which typically sets the present level of fine-tuning at ${\cal O}(1\%)$. However, this result presents a strong dependence on the HE scale. E.g. if the latter is $10^7$~GeV the level of fine-tuning is $\sim$ four times less severe. Finally, the most robust result of Natural SUSY is by far that Higgsinos should be rather light, certainly below 700~GeV for a fine-tuning of ${\cal O}(1\%)$ or milder. Incidentally, this upper bound is not far from 
$\simeq1$~TeV, which is the value required if dark matter is made of Higgsinos.
}

\preprint{IFT-UAM/CSIC-14-068  \\[-8mm]
                    \begin{flushright} 
              ULB-TH/14-11 
                    \end{flushright} }

\begin{document}

\maketitle
\flushbottom

\section{Introduction}

The idea of ``Natural SUSY" has become very popular in the last times, especially as a framework that justifies that e.g. the stops should be light (much lighter than the other squarks), say $m_{\tilde t}\simlt 600$~GeV. This is an attractive scenario since it gives theoretical support to searches for light stops and other particles at the LHC, a hot subject from the theoretical and the experimental points of view.

In a few words, the idea is to lie in a region of the minimal supersymmetric Standard Model (MSSM) parameter-space where the electroweak breaking is not fine-tuned (or not too much fine-tuned). This is reasonable since, as it is usually argued, the main phenomenological virtue of supersymmetry (SUSY) is precisely to avoid the huge fine-tuning associated to the hierarchy problem. 

Then, the main argument is in brief the following: ``Stops produce the main radiative contributions to the Higgs potential, in particular to the Higgs mass-parameter $m^2$. To avoid fine-tunings these contributions should be reasonably small, not much larger than $m^2$ itself. Since they are proportional to the stop masses, the latter cannot be too large". 
Other supersymmetric particles, like gluinos, are also constrained by the same reason; in particular the gluino mass is bounded from above due to its important contribution to the running of the stop masses, which implies a significant 2-loop contribution to $m^2$. In addition, Higgsinos should be light, as their masses are controlled by the $\mu-$parameter, which contributes to $m^2$. These statements sound reasonable and have been often used to quantify the ``naturalness" upper bounds on stop masses, gluino masses, etc. 

Apart from the theoretical arguments, the experiments at the LHC, ATLAS and CMS, performed a large number of SUSY searches, covering a significant part of the parameter space~\cite{Aad:2014pda,Aad:2014wea,Aad:2014bva,ATLAS-CONF-2014-006,Chatrchyan:2013iqa,Chatrchyan:2013xna,Chatrchyan:2014lfa,CMS-PAS-SUS-13-020}. From the point of view of Natural SUSY models, the most interesting bounds are those for stops, gluinos and Higgsinos. The lower limits from direct production of stops reach as far as 650~GeV~\cite{Aad:2014kra}, but they are sensitive to the stop details, in particular the mass difference between the stop and the lightest supersymmetric particle (LSP). Much lighter stops are still allowed by the current experimental bounds once certain conditions on their decays are fulfilled~\cite{Rolbiecki:2013fia,Kim:2014eva,Curtin:2014zua}. Concerning the gluino, the current experimental bounds have a strong dependence on the masses of the light squarks. Assuming  that the stops are the only squarks 
lighter than the gluino (as suggested by the very ``Natural SUSY" rationale), the latter decays through a chain $\tilde{g} \to t\bar{t} \tilde{\chi}^0_1$, and the lower limit reaches $m_{\tilde g}\simlt$ 1.4~TeV~\cite{ATLAS-CONF-2013-061}. Again, with some additional assumptions on the decay chains this limit can be somewhat relaxed. Finally, the $\mu$ parameter is the least constrained at the LHC. Because of the low electroweak production cross-section and the large model dependence, it is entirely possible to have Higgsinos just above the current LEP limit, $\mu \gtrsim 95$~GeV~\cite{Heister:2002mn,Abbiendi:2002vz}. On the other hand, the LEP limit is rather model independent, even if the Higgsino is the LSP with three almost mass degenerate states around 100~GeV.  All these bounds are relevant to establish the present degree of fine-tuning in different SUSY scenarios.

Another important experimental ingredient in connection with Natural SUSY is the physics of the Higgs boson \cite{Aad:2012tfa,Chatrchyan:2012ufa}. In particular, the Higgs mass plays a prominent role in naturalness arguments. According to the most recent analyses,  $m_h = 125.36 \pm 0.41 $~GeV~\cite{Aad:2014aba} (ATLAS) and $m_h = 125.03 \pm 0.30$~GeV~\cite{CMS:2014ega} (CMS).  It is interesting to note that the current measurements are already more accurate than the theoretical predictions, which have a $\sim 2-3$~GeV 
uncertainty~\cite{Feng:2013tvd, Hahn:2013ria, Harlander:2008ju}. 
Furthermore, all observed Higgs properties are remarkably close to the SM predictions~\cite{ATLAS-CONF-2014-009}, which, within the SUSY context, points to the decoupling limit~\cite{Dobado:2000pw}.

In this paper, we revisit the arguments leading to the previous Natural SUSY scenario, showing that some of them are weak or incomplete. In sect.~\ref{sect:rev}, we review the ``standard" Natural SUSY scenario, pointing out some weaknesses in the usual evaluation of its electroweak fine-tuning, i.e. the tuning to get the correct electroweak scale.
We also address the existence of two potential extra fine-tunings that cannot be ignored in the discussion, namely the tuning to get $m_h=m_h^{\rm exp}$ when stops are too light and the tuning to get a large $\tan\beta$. 
In sect.~\ref{sect:EWFT}, we discuss the electroweak fine-tuning of the MSSM, showing its statistical meaning and its generic expression for any theoretical framework. We give in the appendix 
tables and plots that allow to easily evaluate the fine-tuning for any theoretical model within the framework of the MSSM, at any value of the high-energy scale. 
In sect.~\ref{sect:NatBounds}, we describe our method to rigorously extract bounds on the initial (high-energy) parameters and on the supersymmetric spectrum, from the fine-tuning conditions. 
In sect.~\ref{sect:unconstrained}, we apply this method to obtain the numerical values of the various naturalness bounds for the unconstrained MSSM, defined at arbitrary high-energy scale, in a systematic way. 
In sec.~\ref{sect:impactextraFT}, we evaluate the impact of the potential extra fine-tunings mentioned above, discussing also the correlation between soft terms that the experimental Higgs mass imposes in the MSSM and its consequences for the electroweak fine-tuning.
Finally, in sect.~\ref{sect:Summary} we present the summary and conclusions of the paper, outlining the main characteristics of Natural SUSY and their level of robustness against changes in the theoretical framework or the high-energy scale at which the soft parameters appear.

\section{The Natural SUSY scenario. A critical review}\label{sect:rev}

\subsection{The ``standard" Natural SUSY}

Naturalness arguments have been used since long ago \cite{Barbieri:1987fn}  to constrain from above supersymmetric 
masses\footnote{ For a partial  list of references on naturalness in SUSY,  see \cite{deCarlos:1993yy, Anderson:1994dz, Ciafaloni:1996zh, Bhattacharyya:1996dw, Chankowski:1997zh, Barbieri:1998uv, Kane:1998im, Giusti:1998gz, BasteroGil:1999gu, Chan:1997bi, Chacko:2005ra, Choi:2005hd, Nomura:2005qg, Kitano:2005wc, Nomura:2005rj, Lebedev:2005ge, Allanach:2006jc, Perelstein:2007nx, Allanach:2007qk, Cabrera:2008tj, Cassel:2009ps, Kobayashi:2009rn, Lodone:2010kt, Asano:2010ut, Cassel:2011tg, Kitano:2006gv, Giudice:2006sn, Barbieri:2009ev, Feng:1999mn, Feng:1999zg, Romanino:1999ut, Horton:2009ed}  (before LHC )   and 
\cite{ Strumia:2011dv, Akula:2011jx,Antusch:2012gv, 
Martin:2013aha, Evans:2013jna, Hardy:2013ywa, Baer:2013bba, Arvanitaki:2013yja, Kaminska:2013mya, 
Baer:2014ica, Bae:2014fsa, Chakraborti:2014gea, Mustafayev:2014lqa, Fowlie:2014xha, Craig:2013cxa,Antoniadis:2014eta, Kowalska:2014hza, Delgado:2014vha}.
}.
 Already in the LHC era, they were re-visited in ref.~\cite{Papucci:2011wy} to formulate the so-called Natural SUSY scenario. For the purpose of later discussion, we summarize in this subsection the argument of ref.~\cite{Papucci:2011wy}, which have been invoked in many papers.

Assuming that the extra (supersymmetric) Higgs states are heavy enough, the Higgs potential can be written in the Standard Model (SM) way
\begin{eqnarray}
\label{SMpot}
V= m^2|H|^2+\lambda|H|^4 \ ,
\end{eqnarray}
where the SM-like Higgs doublet, $H$, is a linear combination of the two supersymmetric Higgs doublets,  $H\sim \sin\beta H_u + \cos \beta H_d$. Then, the absence of fine-tuning can be expressed as the requirement of not-too-large contributions to the Higss mass parameter, $m^2$. Since the physical Higgs mass is $m_h^2 = 2|m^2|$, a sound measure of the fine-tuning is\footnote{This measure produces similar results to the somewhat standard parametrization of the fine-tuning, see eq.(\ref{BG}) below.} \cite{Kitano:2006gv} 
\be
\label{FT1}
\tilde \Delta=\left|\frac{\delta m^2}{m^2}\right|=\frac{2\delta m^2 }{m_h^2} \ .
\ee
For large $\tan\beta$, the value of $m^2$ is given by $m^2= |\mu|^2 + m_{H_u}^2$, so one immediately notes that both $\mu$ and $m_{H_u}$ should be not-too-large in order to avoid fine-tuning (as has been well-known since many years ago).  For the $\mu-$parameter this implies
\begin{equation}
\mu \simlt 200\GeV\left(\frac{m_{h}}{120\GeV}\right)\left(\frac{\tilde \Delta^{-1}}{20\%}\right)^{-1/2} \ .
\end{equation}
This sets a constraint on Higgsino masses. Constraints for other particles come from the radiative corrections to $m_{H_u}^2$. The most important contribution comes from the stops. Following ref.~\cite{Papucci:2011wy}
\begin{equation}\label{eq:der1}
\delta m_{H_{u}}^{2}|_{\rm stop}=-\frac{3}{8\pi^{2}}y_{t}^{2}\left(m_{Q_{3}}^{2}+ m_{U_{3}}^{2}+|A_{t}|^{2}\right)\log\left(\frac{\Lambda}{\TeV}\right),
\end{equation}
where $\Lambda$ denotes the scale of the transmission of SUSY breaking to the observable sector and  the 1-loop leading-log (LL) approximation was used to integrate the renormalization-group equation (RGE). Then, the above soft parameters $m_{Q_{3}}^{2}$, $m_{U_{3}}^{2}$ and $A_{t}$ are to be understood at low-energy, and thus  they control the stop spectrum. This sets an upper bound on the stop masses.
In particular one has
\begin{equation}
\label{eq:ft-stop}
\sqrt{ \mstl^{2}+\msth^{2}}\lesssim 600\GeV \frac{\sin\beta}{(1+x^{2})^{1/2}}\left(\frac{\log\left(\Lambda/\TeV\right)}{3}\right)^{-1/2}\left(\frac{\tilde \Delta^{-1}}{20\%}\right)^{-1/2}  \, ,
\end{equation}
where $x=A_{t}/\sqrt{\mstl^{2}+\msth^{2}}$. Eq.~(\ref{eq:ft-stop}) imposes a  bound on the lightest stop. Besides the stops, the most important contribution to $m_{H_u}$ is the gluino one, due to its large 1-loop RG correction to the stop masses. Again, in the 1-loop LL approximation used in ref.~\cite{Papucci:2011wy}, one gets
\begin{equation}\label{eq:gluino}
\delta m_{H_{u}}^{2}|_{\rm gluino} \simeq -\frac{2}{\pi^{2}}y_{t}^{2}\left(\frac{\alpha_{s}}{\pi}\right)|M_{3}|^{2}\log^{2}\left(\frac{\Lambda}{\TeV}\right)\, ,
\end{equation}
where $M_{3}$ is the gluino mass. From the previous equation, 
\begin{equation}
M_{3} \lesssim 900\GeV \sin\beta \left(\frac{\log\left(\Lambda/\TeV\right)}{3}\right)^{-1}\left(\frac{m_{h}}{120\GeV}\right)\left(\frac{\tilde \Delta^{-1}}{20\%}\right)^{-1/2} \, . 
\end{equation}

Altogether, the summary of the minimal requirements for a natural SUSY spectrum, as given in ref.~\cite{Papucci:2011wy}, is:
\begin{itemize}
\item two stops and one (left-handed) sbottom, both below $500-700~\GeV$.
\item two Higgsinos, \ie, one chargino and two neutralinos below $200-350~\GeV$. In the absence of other [lighter] chargino/neutralinos, their spectrum is quasi-degenerate.
\item a not too heavy gluino, below $900~\GeV-1.5~\TeV$.
\end{itemize}

In the next subsections we point out the weak points of the above arguments that support the `standard" Natural SUSY scenario. Part of those points have been addressed in the literature after ref.~\cite{Papucci:2011wy} (see ref.~\cite{Feng:2013pwa} for a recent and sound presentation of the  naturalness issue in SUSY and references therein.)

\subsection{The dependence on the initial parameters}
\label{sec:initialcond}

The one-loop LL approximation used to write eqs.(\ref{eq:ft-stop}, \ref{eq:gluino}), from which the naturalness bounds were obtained, is too simplistic in two different aspects. 

First, it is not accurate enough since the top Yukawa-coupling, $y_t$, and the strong coupling, $\alpha_s$, are large and vary a lot along the RG running. As a result, the soft masses evolve greatly and cannot be considered as constant, even as a rough estimate. This effect can be incorporated by integrating numerically the RGE, which corresponds to summing the leading-logs at all orders \cite{Ford:1992mv, Bando:1992np,Bando:1992wy,Einhorn:1982pp}.

Second, and even more important, the physical squark, gluino and electroweakino masses are not initial parameters, but rather a low-energy consequence of the initial parameters at the high-energy scale. This means that one should evaluate the cancellations required among those initial parameters in order to get the correct electroweak scale.
This  entails two complications. First, there is not one-to-one correspondence between the initial parameters and the physical quantities, since the former get mixed along their coupled RGEs. Consequently, it is not possible in general to determine individual upper bounds on the physical masses, not even on the initial parameters. Instead, one should expect to obtain contour-surfaces with equal degree of fine-tuning in the parameter-space  and, similarly, in the ``space" of the possible supersymmetric spectra. The second complication is that the results depend (sometimes critically) on what one considers as initial parameters.

The most dramatic example of the last statements is the dependence of $m_{H_u}^2$ on the stop masses in the constrained MSSM (CMSSM). In the CMSSM one assumes universality of scalar masses at the GUT scale, $M_X$. This is a perfectly reasonable assumption that takes place in well-motivated theoretical scenarios, such as minimal supergravity. Then one has to evaluate the impact of the initial parameters on $m_{H_u}^2$, and see whether or not the requirement of no-fine-tuning implies necessarily light stops. A most relevant analytic study concerning this issue is the well-known work by Feng et al. \cite{Feng:1999mn}, where they studied the focus 
point \cite{Feng:1999mn, Feng:1999zg, Chan:1997bi} region of the CMSSM . In the generic MSSM, the (1-loop) RG evolution of a shift in the initial values of $m_{H_u}^2, m_{U_3}^2, m_{Q_3}^2$ reads 
\begin{equation}
\frac{d}{dt} \left[ \begin{array}{c} \delta m_{H_u}^2
\\ \delta m_{U_3}^2 \\ \delta m_{Q_3}^2 \end{array} \right]
= \frac{y_t^2}{8\pi^2} \left[
\begin{array}{ccc}
3 & 3 & 3 \\
2 & 2 & 2 \\
1 & 1 & 1 \end{array} \right]
\left[ \begin{array}{c} \delta m_{H_u}^2
\\ \delta m_{U_3}^2 \\ \delta m_{Q_3}^2 \end{array} \right] \ , 
\end{equation}
where $t\equiv\ln Q$, with $Q$ the  renormalization-scale, and $y_t$ is the top Yukawa coupling.
Hence, starting with the CMSSM universal condition at $M_X$: $m_{H_u}^2=m_{U_3}^2=m_{Q_3}^2=m_0^2$, one finds
\begin {equation}
\delta m_{H_u}^2  = \frac{\delta m_0^2}{2} \left\{3\  
{\rm exp} \left[ \int_0^t \frac{6y_t^2}{8\pi^2} dt' \right]
- 1  \right\}.
\label{focus}
\end{equation}
Provided $\tan\beta$ is large enough, ${\rm exp}\! \left[ {6\over8\pi^2}
\int_0^t y_t^2 dt' \right] \simeq 1/3$ for the integration between $M_X$ and the electroweak scale,  so the value of $m_{H_u}^2$ depends very little (in the CMSSM) on the initial scalar mass, $m_0$. However, the average stop mass is given by (see eq.(\ref{mtbound1}) below)
\begin{eqnarray}
\label{mstops}
\overline m^2_{\tilde t} \simeq 2.97 M_3^2 + 0.50  m_0^2 + \cdots  \ ,
\end{eqnarray}
where $M_3$ is the gluino mass at $M_X$.
Therefore, if the stops are heavy {\em because} $m_0$ is large, this does {\em not} imply fine-tuning. This is a clear counter-example to the need of having light stops to ensure naturalness.

From the previous discussion it turns out that the most rigorous way to analyze the fine-tuning is to determine the full dependence of the electroweak scale (and other potentially fine-tuned quantities) on the initial parameters, and then derive the regions of constant fine-tuning in the parameter space. 
These regions can be (non-trivially) translated into  constant fine-tuning regions in the space of possible physical spectra.
This goal is enormously simplified if one determines in the first place the analytical dependence of low-energy quantities on the high-energy initial parameters, a task which will be carefully addressed in subsection~\ref{sect:fit}.

\subsection{Fine-tunings left aside} \label{sect:extraFT}

In a MSSM scenario, there are two implicit potential fine-tunings that have to be taken into account to evaluate the global degree of fine-tuning. They stem from the need of having a physical Higgs mass consistent with $m_h^{\rm exp}\simeq 125$~GeV and from the requirement of rather large $\tan\beta$. Let us comment on them in order.

\subsubsection*{Fine-tuning to get $m_h^{\rm exp}\simeq 125$~GeV}

As  is well known, the tree-level Higgs mass in the MSSM is given by $(m_h^2)_{\rm tree-level}=M_Z^2 \cos^22\beta$, so radiative corrections are needed in order to reconcile it with the experimental value. A simplified expression  of such corrections \cite{ Haber:1996fp, Casas:1994us,Carena:1995bx}, useful for the sake of the discussion, is
\begin{equation}\label{eq:der2}
\delta m_{h}^{2}=\frac{3 G_{F}}{\sqrt 2 \pi^{2}}m_{t}^{4}\left(\log\left(\frac{\overline m_{\tilde t}^{2}}{m_{t}^{2}}\right)+\frac{X_{t}^{2}}{\overline m_{\tilde t}^{2}}\left(1-\frac{X_{t}^{2}}{12\overline m_{\tilde t}^{2}}\right)\right)\ +\ \cdots\ ,
\end{equation}
with $\overline m_{\tilde t}$ the average stop mass and $X_{t}=A_{t}-\mu \cot\beta$. The $X_t$-contribution arises from the threshold corrections to the quartic coupling at the stop scale. This correction is maximized for $X_{t}=\sqrt 6 \overline m_{\tilde t}$ ($X_{t}\simeq  2 \overline m_{\tilde t}$ when higher orders are included). Notice that if the threshold correction were not present one would need heavy stops (of about 3~TeV once higher order corrections are included) for large $\tan \beta$ (and much heavier as $\tan\beta$ decreases, see ref. \cite{Cabrera:2011bi,Giudice:2011cg}); which is inconsistent with the requirements of Natural SUSY in its original formulation. However, taking $X_t$ close to the ``maximal" value, it is possible to obtain the correct Higgs mass with rather light stops, even in the $500-700~\GeV$ range; a fact frequently invoked in the literature to reconcile the Higgs mass with Natural SUSY.

On the other side, requiring $X_{t}\sim$ maximal, amounts also to a certain fine-tuning if one needs to lie close to such value with great precision. The precision (and thus the fine-tuning) required depends in turn on the values of $\tan\beta$ and the stop masses. Therefore, when analyzing the naturalness issue one should take into account, besides the fine-tuning associated with the electroweak breaking, the one associated with the precise value required for $X_t$. In subsection~\ref{sect:125} we will discuss the size of this fine-tuning in further detail.

\subsubsection*{Fine-tuning to get large $\tan\beta$}

The value of $\tan\beta \equiv \langle H_u\rangle/\langle H_d\rangle$ is given, at tree level, by
\bea
\label{B}
\frac{2}{\tan\beta}\simeq \sin 2\beta=\frac{2B\mu}{m_{H_d}^2 + m_{H_u}^2+2\mu^2}=\frac{2B\mu}{m_A^2}\ ,
\eea
where $m_A$ is the mass of the pseudoscalar Higgs state; all the quantities above are understood to be evaluated at the low-scale. Clearly, in order to get large $\tan\beta$ one needs small $B\mu$ at low-energy. However, even starting with vanishing $B$ at $M_X$ one gets a large radiative correction due to the RG running. Consequently, very large values of $\tan\beta$ are very fine-tuned\footnote{The existence of this fine-tuning was first observed in ref. \cite{Nelson:1993vc, Hall:1993gn} and has been discussed, from the Bayesian point of view in ref. \cite{Cabrera:2009dm}.}, as they require a cancellation between the initial value of $B$ and the radiative contributions. On the other hand, moderately large values may be non-fine-tuned, depending on the size of the RG contribution to $B\mu$ and the value of $m_A$. Hence, a complete analysis of the MSSM naturalness has to address this potential source of fine-tuning.

\section{The electroweak fine-tuning of the MSSM}\label{sect:EWFT}

In the MSSM, the vacuum expectation value of the Higgs, $v^2/2 = |\langle H_u\rangle|^2 +|\langle H_d\rangle|^2$, is given, at tree-level, by the minimization relation
\bea
\label{mu}
-\frac{1}{8}(g^2+g'^2)v^2 = -\frac{M_Z^2}{2}=\mu^2-\frac{m_{H_d}^2 - m_{H_u}^2\tan^2\beta}{\tan^2\beta-1} \ .
\eea
As is well known, the value of $\tan\beta$ must be rather large, so that the tree-level Higgs mass,  $(m_h^2)_{\rm tree-level}=M_Z^2 \cos^22\beta$, is as large as possible, $\simeq M_Z^2$; otherwise, the radiative corrections needed to reconcile the Higgs mass with its experimental value, would imply gigantic stop masses \cite{Cabrera:2011bi,Giudice:2011cg} (see subsection~\ref{sect:extraFT} above) and thus an extremely fine-tuned scenario. Notice here that the focus-point regime is not useful to cure such fine-tuning since it only works if $\tan\beta$ is rather large and stop masses are not huge.

Therefore, for Natural SUSY the  limit of large $\tan\beta$ is the relevant one. Then, the relation (\ref{mu}) gets simplified
\bea
\label{min}
-\frac{1}{8}(g^2+g'^2)v^2 = -\frac{M_Z^2}{2}=\mu^2+m_{H_u}^2\ .
\eea
The two terms on the  r.h.s have opposite signs and their absolute values are typically much larger than $M_Z^2$, hence the potential fine-tuning associated to the electroweak breaking.

It is well-known that the radiative corrections to the Higgs potential reduce the fine-tuning \cite{deCarlos:1993yy}. This effect can be honestly included taking into account that the effective quartic coupling of the SM-like Higgs runs from its initial value at the SUSY threshold\footnote{A convenient choice of the SUSY-threshold is the average stop mass, since the 1-loop correction to the Higgs potential is dominated by the stop contribution. Hence, choosing $Q_{\rm threshold}\simeq m_{\tilde t}$, the 1-loop correction is minimized and the Higgs potential is well approximated by the tree-level form.}, $\lambda(Q_{\rm threshold})=\frac{1}{8}(g^2+g'^2)$, until its final value at the electroweak scale, $\lambda(Q_{EW})$.
The effect of this running is equivalent to include the radiative contributions to the Higgs quartic coupling in the effective potential, which increase the tree-level Higgs mass, $(m_h^2)_{\rm tree-level}=2\lambda(Q_{\rm threshold})v^2=M_Z^2$, up to the experimental one, $m_h^2=2\lambda(Q_{EW})v^2$. Therefore, replacing $\lambda_{\rm tree-level}$ by the radiatively-corrected quartic coupling is equivalent to replace $M_Z^2 \rightarrow m_h^2$ in eq.(\ref{min}) above, i.e.
\bea
\label{minh}
-\frac{m_h^2}{2}=\mu^2+m_{H_u}^2\ ,
\eea
which is the expression from which we will evaluate the electroweak fine-tuning in the MSSM. As mentioned above, the radiative corrections slightly alleviate this fine-tuning, since $m_h>M_Z$.

\subsection{The measure of the fine-tuning} \label{sect:FTmeasure}

It is a common practice to quantify the amount of fine-tuning using the parametrization first proposed by Ellis et al. \cite{Ellis:1986yg} and Barbieri and Giudice \cite{Barbieri:1987fn}, which in our case reads
\bea
\label{BG}
\frac{\partial m_h^2}{\partial \theta_i} = \Delta_{\theta_i}\frac{ m_h^2}{\theta_i}\ , \ \ \ \ \ \Delta\equiv {\rm Max}\ \left|\Delta_{\theta_i}\right|\ ,
\eea
where $\theta_i$ is an independent parameter that defines the model under consideration and $\Delta_{\theta_i}$ is the fine-tuning parameter associated to it. Typically $\theta_{i}$ are the initial (high-energy) values of the soft terms and the $\mu$ parameter. Nevertheless, for specific scenarios of SUSY breaking and transmission to the observable sector, the initial parameters might be particular theoretical parameters that define the scenario and hence determine the soft terms, e.g. a Goldstino angle in scenarios of moduli-dominated SUSY breaking. We will comment further on this issue in subsection~\ref{sect:genericFT}.

It is worth to briefly comment on the statistical meaning of $\Delta_{\theta_i}$. In ref.~\cite{Ciafaloni:1996zh}  it was argued that (the maximum of all) $|\Delta_{\theta_i}|$ represents the inverse of the probability of a cancellation among terms of a given size to obtain a result which is $|\Delta_{\theta_i}|$ times smaller. This can be intuitively seen as follows. Expanding $m_h^2(\theta_i)$ around a point in the parameter space that gives the desired cancellation, say $\{\theta_i^0\}$,
up to first order in the parameters, one finds that only a small neighborhood $\delta \theta_i\sim \theta^0_i/\Delta_{\theta_i}$ around this point gives a value of $m_h^2$ smaller or equal to the experimental value \cite{Ciafaloni:1996zh}. Therefore, if one assumes that $\theta_i$ could reasonably have taken any value of the order of magnitude of $\theta_i^0$, then only for a small fraction $\left|{\delta \theta_i}/{\theta^0_i}\right| \sim \Delta_{\theta_i}^{-1}$ of this region one gets $m_h^2\lsim (m_h^{\rm exp})^2$, hence the rough probabilistic meaning of $\Delta_{\theta_i}$.
Note that the value of $\Delta$ can be interpreted as the inverse of the $p$-value to get the correct value of $m_h^2$. If $\theta$ is the parameter that  gives the maximum $\Delta$ parameter, then\footnote{Notice that in the particular case when $\theta^0$ minimizes the value of $m_h$, then $\left.\partial m_h/\partial \theta\right|_{\theta=\theta_0}=0$. This lack of sensitivity at first order when $\theta_0$ is close to an stationary point, would seemingly imply no fine-tuning, according to the ``standard criterion". However, from the above discussion, it is clear that in this case the expansion at first order is meaningless; one should start at second order, and then it becomes clear that the fine-tuning is really very high since only when $\theta$ is close to $\theta_0$, one gets $m_h^2\lsim (m_h^{\rm exp})^2$. In other words, the associated $p$-value would be very small.}
\begin{eqnarray}
\label{toy2}
p{\rm -value} \simeq \left|\frac{\delta \theta}{\theta_0}\right| \equiv \Delta^{-1}\ .
\end{eqnarray}
It is noteworthy that for the previous arguments it was implicitly assumed that the possible values of a $\theta_i-$parameter are distributed, with approximately flat probability, in the $[0,\theta^0_i]$ range. In a Bayesian language, the prior on the parameters was assumed to be flat, within the mentioned range. If the assumptions are different (either because the allowed ranges of some parameters are restricted by theoretical consistency or experimental data, or because the priors are not flat), then the probabilistic interpretation has to be consistently modified. These issues become more transparent using a Bayesian approach.

In a Bayesian analysis, the goal is to generate a map of the relative probability of the different regions of the parameter space of the model under consideration (MSSM in our case), using all the available (theoretical and experimental) information. This is the so-called {\em posterior} probability, $p(\theta_i|{\rm data})$, where `data' stands for all the experimental information and $\theta_i$ represent the various parameters of the model. The posterior is given by the Bayes' Theorem
\bea
\label{Bayes}
p(\theta_i|{\rm data})\ =\ p({\rm data}|\theta_i)\ p(\theta_i)\ \frac{1}{p({\rm data})}\ ,
\eea
where $p({\rm data}|\theta_i)$ is the likelihood (sometimes denoted by ${\cal L}$), i.e. the probability density of observing the given data if nature has chosen to be at the $\{\theta_i\}$ point of the parameter space (this is the quantity used in frequentist approaches); $p(\theta_i)$ is the prior, i.e. the ``theoretical" probability density that we assign a priori to the point in the parameter space; and, finally, $p({\rm data})$ is a normalization factor which plays no role unless one wishes to compare different classes of models.

For the sake of concreteness, let us focus on a particular parameter defining the MSSM, namely the $\mu-$parameter\footnote{Of course, one can take here another parameter and the argument goes the same (actually, in some theoretical scenarios $\mu$ may be not an initial parameter). On the other hand, $\mu$ is a convenient choice since it is the parameter usually solved in terms of $M_Z$ in phenomenological analyses.}.
Now, instead of solving $\mu$ in terms of $M_Z$ and the other supersymmetric parameters using the minimization conditions (as usual), one can (actually should) treat $M_Z^{\rm exp}$, i.e. the electroweak scale, as experimental data on a similar footing with the other observables, entering the total likelihood, ${\cal L}$. Approximating the $M_Z$ likelihood as a Dirac delta,
\bea
\label{likelihood}
p({\rm data}|M_1, M_2,\cdots, \mu)\ \simeq\ \delta(M_Z-M_Z^{\rm exp})\ {\cal L}_{\rm rest}\ ,
\eea
where ${\cal L}_{\rm rest}$ is the likelihood associated to all the physical observables except $M_Z$, one can marginalize the $\mu-$parameter
\bea
\label{marg_mu}
&&p(M_1, M_2, \cdots | \ {\rm data} ) = \int d\mu\ p(M_1, M_2, \cdots , \mu | 
{\rm data} )
\nonumber\\ 
&&\hspace{3cm}\propto\ {\cal L}_{\rm rest}  \left|\frac{d\mu}{d M_Z}\right|_{\mu_Z}
p(M_1, M_2, \cdots , \mu_Z)\ ,
\eea
where we have used eqs.~(\ref{Bayes}, \ref{likelihood}). Here $\mu_Z$ is the value of $\mu$ that reproduces $M_Z^{\rm exp}$ for the given values of $\{M_1,M_2,\cdots \}$, and $p(M_1, M_2, \cdots, \mu)$ is the prior in the initial parameters (still undefined). Note that the above Jacobian factor in eq.(\ref{marg_mu}) can be written as\footnote{Notice that the dependence of $M_Z$ on $\mu$ is through eq.(\ref{minh}), which determines the Higgs VEV. Thus $\frac{d M_Z^2}{d \mu}\propto \frac{d m_h^2}{d \mu}$.}
\bea
\label{BG_Bayes}
\left|\frac{d\mu}{d M_Z}\right|_{\mu_Z}\propto 
\left|\frac{\mu}{\Delta_\mu}\right|_{\mu_Z},\
\eea 
where the constant factors are absorbed in the global normalization factor of eq.(\ref{Bayes}). The important point is that the relative probability density of a point in the MSSM parameter space is multiplied by $\Delta_\mu^{-1}$, which is consistent with the above probabilistic interpretation of $\Delta$ \cite{Cabrera:2008tj, Cabrera:2009dm, Ghilencea:2012qk, Fichet:2012sn}. 
Actually, the equivalence is exact if one assumes that the prior in the parameters is factorizable, i.e. $p(M_1, M_2, \cdots , \mu)=p(M_1) p(M_2) \cdots  p(\mu)$, {\em and} $p(\mu_Z)\propto 1/\mu_Z$, so that the numerator in the r.h.s of (\ref{BG_Bayes}) is canceled when plugged in eq.(\ref{marg_mu}). This assumption can be realized in two different ways. First, if $\mu$ has a flat prior with range $\sim [0,\mu_Z]$, then the normalization of the $\mu-$prior goes like $\propto 1/\mu_Z$. This is exactly the kind of implicit assumption discussed above. Alternatively, if $\mu$ has a logarithmically flat prior, then $p(\mu)\propto 1/\mu$, with the same result (this is probably the most sensible prior to adopt 
since it means that all magnitudes of the 
SUSY parameters are equally probable). 

In summary, the standard measure of the fine-tuning (\ref{BG}) is reasonable and can be rigorously justified using Bayesian methods. In consequence, we will use it throughout the paper. Nevertheless, it should be kept in mind that the previous Bayesian analysis also provides the implicit assumptions for its validity. If a particular theoretical model does not fulfill them, the standard criterion is inappropriate and should be consistently modified.

 \subsection{Generic expression for the fine-tuning} \label{sect:genericFT}

Clearly, in order to use the standard measure of the fine-tuning (\ref{BG}) it is necessary to write the r.h.s. of the minimization equation (\ref{minh}) in terms of the initial parameters. This in turn implies to write the low-energy values of $m_{H_u}^2$ and $\mu$ in terms of the initial, high-energy, soft-terms and $\mu-$term (for specific SUSY constructions, these parameters should themselves be expressed in terms of the genuine initial parameters of the model). Low-energy (LE) and high-energy (HE) parameters are related by the RG equations, which normally have to be integrated numerically. However, it is extremely convenient to express this dependence in an exact, analytical way. Fortunately, this can be straightforwardly done, since the dimensional and analytical consistency dictates the form of the dependence,
\begin{eqnarray}
\label{mHu_gen_fit}
m_{H_u}^2(LE)&=&
c_{M_3^2}M_3^2 +c_{M_2^2}M_2^2 +c_{M_1^2}M_1^2 + c_{A_t^2}A_t^2+c_{A_tM_3}A_tM_3 +c_{M_3M_2}M_3M_2+ \cdots 
\nonumber\\
&&\cdots +c_{m_{H_u}^2}m_{H_u}^2 +c_{m_{Q_3}^2} m_{Q_3}^2 +c_{m_{U_3}^2}m_{U_3}^2+\cdots
\\
\mu(LE)&=&c_\mu \mu \ ,
\label{mu_gen_fit}
\end{eqnarray}
where $M_i$ are the $SU(3)\times SU(2)\times U(1)_Y$ gaugino masses, $A_t$ is the top trilinear scalar coupling; and $m_{H_u}, m_{Q_3}, m_{U_3}$ are the masses of the $H_u-$Higgs, the third-generation squark doublet and the stop singlet respectively, all of them understood at the HE scale. The numerical coefficients, $c_{M_3^2}, c_{M_2^2},...$ are obtained by fitting the result of the numerical integration of the RGEs to eqs.(\ref{mHu_gen_fit}, \ref{mu_gen_fit}), a task that we perform carefully in the subsection~\ref{sect:fit}.

The above equations (\ref{mHu_gen_fit}, \ref{mu_gen_fit}) replace the one-loop LL expressions (\ref{eq:der1}, \ref{eq:gluino}) used in the standard Natural-SUSY treatment. If one considers the initial values of the soft parameters and $\mu$ as the independent parameters that define the MSSM, then one can easily extract the associated fine-tuning by applying eq.(\ref{BG}) to (\ref{minh}), and replacing $m_{H_u}^2$ by the expression (\ref{mHu_gen_fit}). Note that the above definition of $\Delta$, eq.(\ref{BG}), is actually not very different from the definition  (\ref{FT1}) used in ref.~\cite{Papucci:2011wy}; actually they are identical for the parameters that enter as a single term in the sum of eq.(\ref{mHu_gen_fit}), e.g. $m_{\tilde U_3}^2$. Nevertheless,  eq.(\ref{BG}) differs from eq.(\ref{FT1}) when the parameter enters in several terms, e.g. $M_3$.\footnote{Indeed, if eq.(\ref{FT1}) was refined to incorporate the $M_3-$dependent contributions to $m^2$, e.g. through their impact in the stop mixing,
 the result would be very similar to that of eq.(\ref{BG}).}
On the other hand, the definition (\ref{BG}), besides being statistically more meaningful, allows to study scenarios where the initial parameters are not soft masses. 

From eqs.(\ref{minh}, \ref{mHu_gen_fit}, \ref{mu_gen_fit})) it is easy to derive the $\Delta-$parameters (\ref{BG}) for any MSSM scenario. 
A common practice is to consider the (HE) soft terms and the $\mu-$term as the independent parameters, say 
\begin{eqnarray}
\label{Thetas}
\Theta_\alpha=\left\{\mu, M_3,M_2,M_1,A_t, m_{H_u}^2, m_{H_d}^2, m_{U_3}^2, m_{Q_3}^2, \cdots\right\}, 
\end{eqnarray}
which is equivalent to the so-called ``Unconstrained MSSM"\footnote{{ The name ``Unconstrained MSSM" could be a bit misleading in this context, since it would seem to imply that one is not doing any assumptions about the soft terms. But there is in fact an assumption, namely that they are not correlated. Note in particular that  although the parameter space of the Unconstrained MSSM includes any MSSM, e.g. the ``Constrained MSSM", the calculation of the fine-tuning for the latter requires to take into account a specific correlation between various soft-terms. Still, we are showing in this section that the results for the Unconstrained MSSM allow to easily evaluate the fine-tuning in any other MSSM scenario.}} .Then one easily computes $\Delta_{\Theta_\alpha}$
\begin{eqnarray}
\label{BGTheta}
\Delta_{\Theta_\alpha}=\frac{\Theta_\alpha}{m_h^2}\frac{\partial m_h^2}{\partial \Theta_\alpha} = -2\frac{\Theta_\alpha}{m_h^2}\frac{\partial m_{H_u}^2}{\partial \Theta_\alpha}  \ .
\end{eqnarray}
E.g. $\Delta_{M_3}$ is given by
\begin{eqnarray}
\label{deltaBGM3}
\Delta_{M_3} = -2\frac{M_3}{m_h^2}
\left(2c_{{M_3^2}} M_3+c_{A_tM_3}A_t+c_{M_3M_2}M_2+\cdots
\right)\ .
\end{eqnarray}
The identification $\frac{\partial m_h^2}{\partial \Theta_\alpha}\simeq -2\frac{\partial m_{H_u}^2}{\partial \Theta_\alpha}$ in eq.(\ref{BGTheta}) comes from eq.(\ref{minh}) and thus is valid for all the parameters except $\mu$, for which we simply have
\begin{eqnarray}
\label{BGmu}
\Delta_{\mu}=\frac{\mu}{m_h^2}\frac{\partial m_h^2}{\partial \mu} 
=-4 c_\mu^2 \df{\mu^2}{m_h^2} =  -4 \left(\df{\mu(LE)}{m_h^2}\right)^2 \ .
 \end{eqnarray} 
 Besides, the term proportional to $m_{H_d}^2$  in eq.(\ref{mu}), which was subsequently neglected, can give relatively important contributions to $\Delta_{m_{H_d}^2}$ if $\tan\beta$ is not too large ($\simlt 10$), namely
\begin{eqnarray}
\label{BGmHd}
\Delta_{m_{H_d}^2}\simeq -2\frac{m_{H_d}^2}{m_h^2}\left(c_{m_{H_d}^2} - c'_{m_{H_d}^2}\ {({\tan^2\beta-1})^{-1}\ }\ \right)\ ,
 \end{eqnarray}
where $c'_{m_{H_d}^2}\simeq 1$ denotes the $c-$coefficient of $m_{H_d}^2$ in the expression of the LE value of $m_{H_d}^2$ itself, see table \ref{tab:mhiggssq} in the appendix. In any case, the contribution of $m_{H_d}^2$ to the fine-tuning is always marginal.

Note that for any other theoretical scenario, the $\Delta$s associated with the 
genuine initial parameters, say $\theta_i$, can be written in terms of $\Delta_{\Theta_\alpha}$ 
using the chain rule
\begin{eqnarray}
\label{chain}
\Delta_{\theta_i}\equiv\frac{\partial\ln m_h^2}{\partial \ln\theta_i}
=\sum_\alpha\Delta_{\Theta_\alpha} \frac{\partial \ln \Theta_\alpha}   {\partial \ln\theta_i}
=
\frac{\theta_i}{m_h^2}
\sum_\alpha\frac{\partial m_h^2}{\partial \Theta_\alpha} 
 \frac{\partial\Theta_\alpha}   {\partial \theta_i}\ .
\end{eqnarray}
Finally, in order to obtain fine-tuning bounds on the parameters of the model we demand $\left|\Delta_{\theta_i}\right| \simlt \Delta^{\rm max}$, where 
$\Delta^{\rm max}$ is the maximum amount of fine-tuning one is willing to accept. E.g. 
\begin{eqnarray}
\label{Deltamax}
\Delta^{\rm max}=100 \ ,
\end{eqnarray}
represents a fine-tuning of $\sim 1\%$.

\subsection{The fit to  the low-energy quantities} \label{sect:fit}

Fits of the kind of eq.(\ref{mHu_gen_fit}) can be found in the literature, see e.g.\cite{Abe:2007kf,Martin:2007gf}. 
However, though useful, they should be refined in several ways in order to perform a precise fine-tuning analysis. 
The most important improvement is a careful treatment of the various threshold scales. 
In particular, the initial MSSM parameters (i.e. the soft terms and the $\mu-$parameter) are defined at a high-energy (HE) scale, which is usually identified as $M_X$, i.e. 
the scale at which the gauge couplings unify. Although this is a reasonable assumption, it is convenient to consider the HE scale as an unknown; e.g. in gauge-mediated scenarios it can be in principle any scale. 
The low-energy (LE) scale at which one sets the SUSY threshold and the supersymmetric spectrum is computed, is also model-dependent. 
A reasonable choice is to take $M_{\rm LE}$ as the averaged stop masses. As discussed above, at this scale the 1-loop corrections to the effective potential are minimized, 
so that the potential is well approximated by the tree-level expression; thus eq.(\ref{mHu_gen_fit}) should be understood at this scale. 
Nevertheless, in many fits from the literature $M_{\rm LE}$ is identified with $M_Z$. Finally, some parameters are inputs at $M_Z$, e.g. the gauge couplings, while
others, like the soft $B-$parameter (the coefficient of the bilinear Higgs coupling), have to be evaluated in order to reproduce the correct electroweak breaking with the value of $\tan\beta$ chosen. 
Similarly, the value of the top Yukawa-coupling has to be settled at high energy in such a way that it reproduces the value of the top mass at the electroweak scale (which is below the LE scale). 
All this requires to divide the RG-running into two segments, $[M_{\rm EW},\  M_{\rm LE}]$ and $[M_{\rm LE},\  M_{\rm HE}]$. 
Besides this refinement, we have integrated the RG-equations at two-loop order, using {\tt SARAH 4.1.0} \cite{Staub:2013tta}.

The results of the fits for all the LE quantities for $\tan\beta = 10$ and $M_{\rm HE}=M_X$
are given in the appendix, Tables \ref{tab:mhiggssq}, \ref{tab:msquarkthsq}, \ref{tab:msquarkstsq}, \ref{tab:msqslept}, \ref{tab:gauginos}, \ref{tab:trilinears}, 
and \ref{tab:mu-Bmu} . 
The value quoted for each $c-$coefficient has been evaluated at $M_{\rm LE} = 1$~TeV. The dependence of the $c-$coefficients on $M_{\rm LE}$ is logarithmic and can be well approximated by
\begin{eqnarray}
\label{scaledep}
c_i(M_{\rm LE}) \simeq c_i(1\ {\rm~TeV}) + b_i\ln\frac{M_{\rm LE}}{1\ {\rm~TeV}}\ .
\end{eqnarray}
The value of the $b_i$ coefficients is also given in Tables \ref{tab:mhiggssq}--\ref{tab:mu-Bmu} (for $M_{\rm HE}=M_X$). Certainly, the value of $M_{\rm LE}\sim \overline{m_{\tilde t}}$ is itself a (complicated) function of the initial soft parameters. 
Nevertheless, it is typically dominated by the (RG) gluino contribution,  $M_{\rm LE}\sim \overline{m_{\tilde t}}\sim\sqrt{3}|M_3|$ for $M_{\rm HE}=M_X$. This represents an additional dependence of $m_{H_u}^2$ on $M_3$, which should be taken into account when computing $\Delta_{M_3}$. Actually, this effect diminishes the fine-tuning associated to $M_3$ (which is among the most important ones) because the impact of an increase of $M_3$ in the value of $m_{H_u}^2$ becomes (slightly) compensated by the increase of the LE scale and the consequent
decrease of the $c_{M_3^2}$ coefficient in eq.(\ref{mHu_gen_fit}). We have incorporated this fact in the computations of the fine-tuning. 

 Let us now turn to the dependence of the fine-tuning on the high-energy scale, $M_{\rm HE}$. The absolute values of all the $c-$coefficients in the fits decrease with $M_{\rm HE}$, 
 except perhaps the coefficient that multiplies the parameter under consideration (e.g. $c_{m_{H_u}^2}$ in eq.(\ref{mHu_gen_fit})). 
 In the limit $M_{\rm HE}\rightarrow M_{\rm LE}$ the latter becomes 1, and the others go to zero. Obviously, the fine-tuning decreases as $M_{\rm HE}$ decreases. 
 The actual dependence of the $c-$coefficients on $M_{\rm HE}$ has to do with the loop-order at which it arises. 
 If it does at one-loop, the dependence is logarithmic-like, e.g. for $c_{M_2^2}$ in eq.(\ref{mHu_gen_fit}); if it does at two-loop, 
 the dependence goes like $\sim (\log M_{\rm HE})^2$, e.g. for $c_{M_3^2}$. These dependences are shown in Figs. \ref{fig:mHusq}, \ref{fig:mQ3sq}, \ref{fig:mU3sq}, \ref{fig:Mgauginos} and \ref{fig:Bmu}.

In summary, with the help of Tables \ref{tab:mhiggssq}--\ref{tab:mu-Bmu} and Figs \ref{fig:mHusq}--\ref{fig:Bmu}  it is straightforward to evaluate the fine-tuning parameters of any MSSM scenario.

\section{The naturalness bounds}\label{sect:NatBounds}

\subsection{Bounds on the initial (high-energy) parameters} \label{sect:HEbounds}

Let us explore further the size and structure of the fine-tuning, and the corresponding bounds on the initial parameters, in the unconstrained MSSM, i.e. taking as initial parameters the HE values of the soft terms and the $\mu$-term: $\Theta_\alpha=\left\{\mu, M_3,M_2,M_1,A_t, m_{H_u}^2, m_{H_d}^2, m_{U_3}^2, m_{Q_3}^2, \cdots\right\}$.
This is interesting by itself, and, as discussed above, it can be considered as the first step to compute the fine-tuning in any theoretical scenario. For any of those parameters we demand
\begin{eqnarray}
\label{FTbound}
\left|\Delta_{\Theta_\alpha}\right| \simlt \Delta^{\rm max}\ ,
\end{eqnarray}
where $\Delta_{\Theta_\alpha}$ are given by eq.(\ref{BGTheta}).
Now, for the parameters that appear just once in  eqs.(\ref{mHu_gen_fit}, \ref{mu_gen_fit}) the corresponding naturalness bound (\ref{FTbound}) is trivial and has the form of an upper limit on the parameter size. For dimensional reasons this is exactly the case for dimension-two parameters in mass units, e.g. for  the squared stop masses 
\begin{eqnarray}
\label{BGmQ3}
\left|\Delta_{m_{Q_3}^2}\right|=  \left|-2\frac{m_{Q_3}^2}{m_h^2}\  c_{m_{Q_3}^2}\right|\simlt \Delta^{\rm max} \ \ \longrightarrow\ \ \ 
m_{Q_3}^2\simlt 1.36\ \Delta^{\rm max}\  m_h^2 
\end{eqnarray}
\begin{eqnarray}
\label{BGmU3}
\left|\Delta_{m_{U_3}^2}\right|=  \left|-2\frac{m_{U_3}^2}{m_h^2}\  c_{m_{U_3}^2}\right|\simlt \Delta^{\rm max} \ \ \longrightarrow\ \ \ 
m_{U_3}^2\simlt  1.72\ \Delta^{\rm max}\  m_h^2 \ ,
\end{eqnarray}
where we have plugged ,  $c_{m_{Q_3}}=-0.367$,  $c_{m_{U_3}}=-0.29$, which correspond to  $M_{\rm HE}=M_X$ and $M_{\rm LE}= 1$~TeV, see Table~\ref{tab:mhiggssq}. 
For $\Delta^{\rm max}=100$, we get  
$m_{Q_3}\simlt 1.46$~TeV, $m_{U_3}\simlt 1.64$~TeV, substantially higher than the usual quoted bounds \cite{Feng:2013pwa}. This is mainly due to the refined RG analysis and the use of the radiatively upgraded expression eq.(\ref{minh}),
 rather than eq.(\ref{min}), to evaluate the fine-tuning. We stress that these are the bounds on the high-energy soft masses, the bounds on the physical masses will be worked out in subsection~\ref{sect:boundsspectra}.
The naturalness bounds for the other (HE) dimension-two parameters ($m_{D_{3}}^2$, $m_{Q_{1,2}}^2$, $m_{U_{1,2}}^2$, $m_{D_{1,2}}^2$, $m_{L_3}^2$, ...) have a form similar to eqs.(\ref{BGmQ3}, \ref{BGmU3}) and are also higher than usually quoted. Due to its peculiar RGE, this is also the case of the $\mu-$parameter, see eq.(\ref{BGmu}).

\begin{figure}[ht]
\centering 
\includegraphics[width=.5\linewidth]{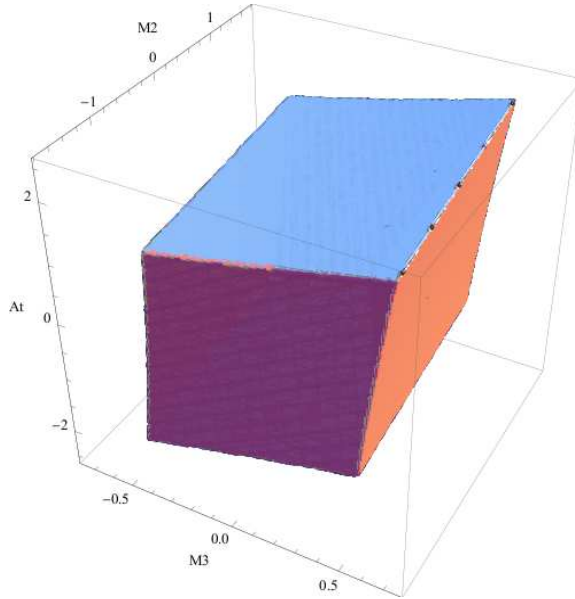}
\caption{
Region in the $\{M_2, M_3, A_t\}$ space that fulfills eq. eqs.(\ref{BGM3}--\ref{BGAt})  for  $\Delta^{\rm max}=100$  (axes units:~TeVs).  For other values, a 
$\sqrt {\Delta^{\rm max}/100 }$ scaling factor has to be applied. }
\label{fig:paralelogram}
\end{figure}

On the other hand, for dimension-one parameters  (except $\mu$) the naturalness bounds (\ref{FTbound}) appear mixed. In particular, this is the case for the bounds associated to $M_3, M_2, A_t$. From eqs.(\ref{BGTheta}) and (\ref{mHu_gen_fit})
\begin{eqnarray}
\label{BGM3}
\left|\Delta_{M_3}\right|&=&\frac{1}{m_h^2}\left|6.41M_3^2 -0.57A_tM_3+0.27M_3M_2\right| \simlt \Delta^{\rm max}
\\
\left|\Delta_{M_2}\right|&=&\frac{1}{m_h^2}\left|-0.81M_2^2 -0.14A_tM_2+0.27M_3M_2\right| \simlt \Delta^{\rm max}
\label{BGM2}\\
\left|\Delta_{A_t}\right|&=&\frac{1}{m_h^2}\left|0.44A_t^2 -0.57A_tM_3-0.14A_tM_2\right| \simlt \Delta^{\rm max} \ ,
\label{BGAt}
\end{eqnarray}
where, again, we have plugged the values of the $c-$coefficients corresponding to $M_{\rm HE}=M_X$ and $M_{\rm LE}= 1$~TeV. Other parameters, like $M_1, A_b$, get also mixed with them in the bounds, but their coefficients are much smaller, so we have neglected them. We show in 
 figure~\ref{fig:paralelogram} the region in the  $\{M_2, M_3, A_t\}$ space that fulfills  the inequalities for $\Delta^{\rm max} = 100$.
The figure is close to a prism.  Their faces are given by the following  approximate  solution to eqs.(\ref{BGM3}--\ref{BGAt})
\begin{eqnarray}
\label{M3max}
M_3^{\rm max}\ \simeq\ \pm\ m_h \sqrt{\frac{\Delta^{\rm max}}{6.41}} + \frac{1}{12.82}(0.57A_t-0.27M_2)
\end{eqnarray}
\begin{eqnarray}
\label{M2max}
M_2^{\rm max}\ \simeq\  \pm\ m_h \sqrt{\frac{\Delta^{\rm max}}{0.81}} + \frac{1}{1.62}(0.27M_3-0.14A_t)
\end{eqnarray}
\begin{eqnarray}
\label{Atmax}
A_t^{\rm max}\ \simeq\ \pm\ m_h \sqrt{\frac{\Delta^{\rm max}}{0.44}} + \frac{1}{0.88}(0.57M_3+0.14M_2)\ ,
\end{eqnarray}
where the superscript ``max" denotes the, positive and negative, values of the parameter that saturate inequalities (\ref{BGM3}--\ref{BGAt}).  
Thus eqs.(\ref{M3max}--\ref{Atmax}) represent the naturalness bounds to $M_3, M_2, A_t$. Each individual bound depends on the values of the other parameters due to the presence of the mixed terms. Depending on the relative signs of the soft terms, the bounds can be larger or smaller than those obtained when neglecting the mixed terms. However, the presence of the latter
stretches each individual {\em absolute} upper bound in a non-negligible way, by doing an appropriate choice of the other soft terms (compatible with their own fine-tuning condition).

A generic, approximate, expression for the absolute upper bound on a  dimension-one  parameter, i.e.  ${\cal M}_i$ $( {\cal M}_i  = M_3, M_2, M_1, A_t, A_b ... )$ can be obtained by replacing the other dimension-one parameters, ${\cal M}_{j\neq i}$, by the values that saturate their zeroth-order fine-tuning bounds, $  \pm {\cal M}_j^{\rm max}\simeq  m_h\sqrt {\Delta^{\rm max}/4|c_{{\cal M}_j^2}| }$, with the appropriate sign; namely
\begin{equation}
|{\cal M}_i |  <   \frac{m_h}{2} \sqrt{ \frac{\Delta^{\rm max}}{ | c_{{\cal M}_i^2 } |}}  
                 \left(  
                         1 + \sum_{j \neq i} \frac{1}{4}  \frac { | c_{{\cal M}_i {\cal M}_j} |} { \sqrt{ | c_{{\cal M}_i^2 } c_{{\cal M}_j^2 } |} }                        
                  \right) \ .
\label{IneqGen}                  
\end{equation}

In practice, in order to obtain the absolute upper bounds on $M_3, M_2, A_t$ we have ignored the presence of additional parameters ($M_1, A_b, \cdots$) in (\ref{IneqGen}). Its inclusion would stretch even further the absolute bounds, but quite slightly and artificially since this would imply a certain conspiracy between soft parameters.
As a matter of fact, even playing just with the three  parameters which show a sizeable correlation, i.e. $\{M_3, M_2, A_t\}$, implies a certain degree of conspiracy to get the maximum value quoted in (\ref{IneqGen}). 
This means that the bound (\ref{IneqGen}) is conservative. A more restrictive and rigorous bound can be obtained by  demanding that the addition in quadrature of the $\Delta_i$ parameters never exceeds the reference value, $\Delta^{\rm max}$. In any case, since the fine-tuning conditions of  $M_3, M_2, A_t$ are correlated, as shown in eqs.~(\ref{BGM3}--\ref{BGAt}), the most meaningful approach is to determine the regions of the parameter space simultaneously consistent with all the fine-tuning conditions. This will be done in detail in section \ref{sect:Hmass} below.
The numerical modification of eqs.(\ref{BGM3}--\ref{Atmax}) for different values of $M_{\rm LE}$, $M_{\rm HE}$ can be  straightforwardly obtained from Table \ref{tab:mhiggssq} and Figure \ref{fig:mHusq}.

Choosing $\Delta^{\rm max}=100$, eqs.(\ref{M3max}--\ref{Atmax}) give $|M_3| \simlt 610$~GeV, $|M_2| \simlt 1630$~GeV, $|A_t| \simlt 2430$~GeV. The limit on $M_3$ is similar to the one found by Feng \cite{Feng:2013pwa}, although this is in part a coincidence. In ref.~\cite{Feng:2013pwa} it was chosen $M_3^2$, rather than $M_3$, as an independent parameter; which reduces the associated $\Delta_{M_3}$ by a factor of 2. So, their bound on $M_3$ was increased (quite artificially in our opinion) by $\sqrt{2}$. On the other hand, in ref.~\cite{Feng:2013pwa} the RG running was not done in two steps, but simply running all the way from $M_X$ till $M_Z$. 
Furthermore, they did not consider the mixed terms of eq.(\ref{BGM3}). And finally they used eq.(\ref{min}) instead of eq.(\ref{minh}) to evaluate the fine-tuning. It turns out that, all together, these three approximations increase the estimate of the fine tuning, thus decreasing the upper bound on $M_3$ by a factor which happens to be $\sim 1/\sqrt{2}$.

Actually, for the particular case of the $M_3-$parameter this is not the end of the story. As discussed in subsection~3.2, the $c_{M_3^2}$ coefficient has a dependence on $M_{\rm LE}$ approximately given by eq.(\ref{scaledep}). Since $M_{\rm LE}\simeq \overline{m_{\tilde t}}$ and typically $\overline{m}^2_{\tilde t}\simeq \frac{1}{2}(c_{M_3^2}^{(Q_3)}+c_{M_3^2}^{(U_3)})^2M_3^2$, where $c_{M_3^2}^{(Q_3)}$, $c_{M_3^2}^{(U_3)}$ are the coefficients of $M_3^2$ in the LE expression of $m^2_{Q_3}$, $m^2_{U_3}$ (given in table \ref{tab:msquarkthsq} and Figs.~\ref{fig:mQ3sq},  \ref{fig:mU3sq} for any HE scale), one has an additional contribution to the computation of $\Delta_{M_3}$ in eq.(\ref{BGTheta}). 
The corresponding correction to $M_3^{\rm max}$ can be estimated by expanding the new inequality around the previous value of  $M_3^{\rm max}$. We find
\begin{eqnarray}
\label{deltaM3}
\delta M_3^{\rm max} \simeq  \frac{1}{2}\frac{b_{M_3}}{|c_{M_3^2}|}
\left( \frac{   \sqrt{\frac{1}{2}\left(c_{M_3^2}^{(Q_3)}+c_{M_3^2}^{(U_3)}\right)}\ M_3^{\rm max}}{1\  {\rm~TeV}}
-\frac12\right)\ M_3^{\rm max} \ ,
\end{eqnarray}
where we have neglected subdominant terms\footnote{Note that this correction is applicable as long as $M_{\rm HE}$ is large ($\simgt 10^{10}$~GeV); otherwise, it is quite small, the stop mass is not determined anymore by $M_3$.}.
For $M_{\rm HE}=M_X$ and  $M_{\rm LE}= 1\  {\rm~TeV} $  one has $c_{M_3^2}\simeq -1.6$, $c_{M_3^2}^{(Q_3)}+c_{M_3^2}^{(U_3)}\simeq 6$, so the previous correction becomes
\begin{eqnarray}
\label{deltaM3_MX}
\delta M_3^{\rm max} \simeq \frac{2.06 M_3^{\rm max}-0.6 {\rm~TeV}}{10 {\rm~TeV}}\ M_3^{\rm max} \ .
\end{eqnarray}
This increases further $M_3^{\rm max}$ from 610~GeV to $\sim$ 660~GeV, i.e. $m_{\tilde g}\simlt 1440$~GeV which is about the present experimental lower limit on the gluino mass. Recall that this bound has been obtained assuming $\Delta^{\rm max}=100$, thus we conclude that the unconstrained MSSM is fine-tuned at about 1\%. We emphasize that these results have been obtained  in the framework of the "unconstrained MSSM", so that $M_3,M_2,A_t$ are treated as independent, non-theoretically-correlated, parameters; and under the assumption $M_{\rm HE}=M_X$.

\subsection{Correlations between the soft terms}

Using the chain rule (\ref{chain}) one can easily evaluate the fine-tuning bounds when the initial soft terms are related in any way determined by the theoretical framework chosen. For instance, it is reasonable to assume that the soft masses at HE come from the same source, and therefore they are related, even if they are not equal. E.g. suppose that at HE

\begin{eqnarray}
\label{relations1}
\left\{m_{H_u}^2, m_{Q_3}^2, m_{U_3}^2\right\} = \left\{a_{H_u}, a_{Q_3}, a_{U_3}\right\} m_0^2 \ .
\end{eqnarray}
Then, plugging eq.(\ref{mHu_gen_fit}) into eq.(\ref{chain}) one immediately derives the fine-tuning condition for $m_0^2$
\begin{eqnarray}
\label{Deltam0}
\left|\Delta_{m_0^2}\right| = \left| -2 \frac{m_0^2}{m_h^2}\left(c_{m_{H_u}^2}a_{H_u} + c_{m_{Q_3}^2}a_{Q_3} + c_{m_{U_3}^2}a_{U_3}
\right)\right|\ \simlt\ \Delta^{\rm max}\ ,
\end{eqnarray}
which entails an upper bound on $m_0^2$, and hence on the stop masses at high energy. E.g.
\begin{eqnarray}
\label{mtRbound}
m_{U_3}^2\simlt \frac{1}{2}\left|\frac{\Delta^{\rm max}}{-0.29+0.631\frac{a_{H_u}}{a_{U_3}}-0.367 \frac{a_{Q_3}}{a_{U_3}}}\right| m_h^2\ ,
\end{eqnarray}
where we have used the $c-$coefficients corresponding to $M_{\rm LE}=1$~TeV, $M_{\rm HE}=M_X$ (Table~\ref{tab:mhiggssq}). This bound can be compared with the bound for the unconstrained MSSM, eq.(\ref{BGmU3}). 
Depending on the relative values between the $a${\scriptsize s} , the bound on 
$m_{U_3}^2$ gets increased (the usual case) or decreased. For the universal case, $a_{H_u}= a_{Q_3}= a_{U_3}$, one gets 
$m_{U_3}\simlt \sqrt{\Delta^{\rm max}}\ 550\  {\rm~GeV}$, which allows for quite heavy stops with very little fine-tuning.

The same game can be played with the gaugino masses and the trilinear couplings. E.g. suppose that
\begin{eqnarray}
\label{relations2}
\left\{M_1,M_2,M_3, A_t\right\} = \left\{a_1, a_2, a_3, a_t\right\} M_{1/2} \ .
\end{eqnarray}
Then, the fine-tuning condition for $M_{1/2}$ reads $\left|\Delta_{M_{1/2}}\right|\simlt\ \Delta^{\rm max}$, with
\begin{eqnarray}
\label{DeltaM12}
\Delta_{M_{1/2}}= -4 \frac{M_{1/2}^2}{m_h^2}\left(c_{M_3^2}a_{3}^2 + c_{M_2^2}a_{2}^2 +c_{A_t^2}a_{t}^2+ c_{M_3M_2}a_3a_2+ c_{M_3A_t}a_3a_t+ c_{M_2A_t}a_2a_t
\right)\ .
\end{eqnarray}
E.g. the bound on $M_3$ becomes
\begin{eqnarray}
\label{m1/2bound}
M_{3}^2\ \simlt\ \frac{a_3^2}{4} \left|\frac{\Delta^{\rm max}}{1.6a_3^2-0.203a_2^2+ 0.109 a_t^2+
0.134a_3a_2 -0.285 a_3a_t -0.068a_2a_t}\right| m_h^2 \ ,
\end{eqnarray}
where, once more, we have used the $c-$coefficients corresponding to $M_{\rm LE}=1$~TeV, $M_{\rm HE}=M_X$.
For the universal case, $a_3=a_2=a_t$, the bound on $M_3$ becomes similar to that of the unconstrained MSSM. However, for other combinations the bound can be much larger. E.g. for $\frac{a_2}{a_3} = 3.16, -2.50$ and $a_t=0$ the 
denominator would cancel.\footnote{See \cite{Antusch:2012gv, Martin:2013aha,Kaminska:2013mya} for some studies about  non-universal gaugino masses and fine-tuning.} This represents a different kind of focus-point, in this case for gauginos.

Other correlations between the soft parameters and the appearance of alternative focus-point regimes can be explored in a similar way starting at any HE scale,  
by using the tables and figures of the appendix. See refs.~\cite{Kowalska:2014hza, Delgado:2014vha} for recent work on this subject.

\subsection{Bounds on the supersymmetric spectrum} \label{sect:boundsspectra}

So far, in this section we have explained in detail how to extract the naturalness limits on the initial (HE) soft terms and $\mu-$term in generic MSSM scenarios. The next step is to translate those bounds into limits on the physical supersymmetric spectrum. Therefore, one has to go back from the high-energy scale to the low-energy one, using the RG equations. Once more, this can be immediately done using the analytical expressions discussed in subsection~\ref{sect:fit} and the appendix for any value of the HE and the LE scales.

Unfortunately, there is no a one-to-one correspondence between the physical masses, and the soft-parameters and $\mu-$term at high-energy. The only approximate exception are the gaugino and Higgsino masses. Namely, from Tables  \ref{tab:gauginos}, \ref{tab:mu-Bmu}
\begin{eqnarray}
\label{Mgauginoshiggsinos}
&&M_{\tilde g} \simeq M_3(M_{\rm LE})\simeq 2.22 M_3 
\nonumber\\
&&M_{\tilde W} \simeq M_2(M_{\rm LE})\simeq 0.81 M_2
\nonumber\\
&&M_{\tilde B} \simeq M_1(M_{\rm LE})\simeq 0.43 M_1
\nonumber\\
&&M_{\tilde H} \simeq \mu(M_{\rm LE})\simeq  1.002 \mu\ ,
\end{eqnarray}
where the above numbers correspond to $M_{\rm LE}=1$~TeV, $M_{\rm HE}=M_X$.
Of course, these are not yet the physical masses, except, approximately, for the gluino. 
For a more precise calculation of the physical (pole) gluino mass, we must incorporate radiative corrections which depend on the 
size of the squark masses and that can be rather significant for more than one squark generation with $m_{\tilde q} \gg M_3$ \cite{Martin:1993yx}.  
The other gauginos and the Higgsinos get mixed in the chargino and neutralino mass matrices.  
However, since we are considering upper limits on these masses, the mixing entries in those matrices are subdominant and do not appreciably affect the bounds. 
On the other hand, as discussed in subsection~\ref{sect:HEbounds}, the naturalness limits on (the HE values of)$M_3$, $M_2$ are more involved than for other parameters, since the respective fine-tuning inequalities get mixed with each other and with $A_t$. Using the ($M_{\rm LE}=1$~TeV, $M_{\rm HE}=M_X$) limits on $M_3, M_2, M_1, \mu$ obtained for the unconstrained MSSM (see sects.~\ref{sect:HEbounds} and \ref{sect:unconstrained}) one gets 
$M_{\tilde g}  \simlt  1440 $~GeV, 
$M_{\tilde W} \simlt  1300 $~GeV,
$M_{\tilde B} \simlt   3370   $~GeV and  
$M_{\tilde H} \simlt    627     $~GeV.

On the contrary, the physical masses of the sparticles, $m_{\tilde t_1}^2$, $m_{\tilde t_2}^2$, $m_{Q_{1,2}}^2$ $m_{U_{1,2}}^2$, $m_{D_{1,2}}^2$, $m_{H^\pm}^2$, etc., are non-trivial combinations of the various initial soft terms and products of them. The case of the stops is particularly important, since it is a common assumption that Natural SUSY demands light stops. E.g. using $M_{\rm LE}=1$~TeV, $M_{\rm HE}=M_X$, we see from table~\ref{tab:mhiggssq} that the values of $m_{Q_3}^2$, $m_{U_3}^2$ at LE are given by:
\begin{eqnarray}
\label{mtLR_fit}
\hspace{-0.8cm} m_{Q_3}^2(M_{\rm LE})&=&   3.191M_3^2 +0.333M_2^2  +0.871 m_{\tilde Q_3}^2   -0.095m_{\tilde U_3}^2  -0.118m_{H_u}^2+0.072 A_tM_3+\cdots
\nonumber\\
\hspace{-0.8cm} m_{U_3}^2(M_{\rm LE})&=&  2.754M_3^2  -0.151 M_2^2  -0.192 m_{\tilde Q_3}^2    +0.706 m_{\tilde U_3}^2  -0.189 m_{H_u}^2 +0.159 A_tM_3+\cdots
\end{eqnarray}
These are not yet the physical stop masses. One has to take into account the top contribution, $m_t^2$, and the off-diagonal entries in the stop mass matrix, $\sim m_t X_t$ where $X_t = A_t+\mu \cot \beta\simeq A_t$. Finally, one has to extract the mass eigenvalues, $ m_{\tilde t_1}^2$ and $ m_{\tilde t_2}^2$. A representative, and easier to calculate, quantity is the average stop mass, 
\begin{eqnarray}
\label{mtbound1}
\overline m_{\tilde t}^2&\equiv& \frac{1}{2}(m_{\tilde t_1}^2 +m_{\tilde t_2}^2)=\frac{1}{2}(m_{Q_3}^2(M_{\rm LE})+m_{U_3}^2(M_{\rm LE}))+m_t^2
\nonumber\\
&\simeq&  (2.972 M_3^2+0.339  m_{Q_3}^2+0.305m_{U_3}^2 + 0.091M_2^2 - 0.154m_{H_u}^2 \cdots) +m_t^2 \ .
\end{eqnarray}
The average stop mass is also an important quantity to evaluate the threshold correction to the Higgs mass, and thus it plays an important role in the evaluation of the potential fine-tuning associated to it, see eq.~(\ref{eq:der2}) and subsection~\ref{sect:125}. Setting $M_3$, $m_{Q_3}$, $m_{U_3}$ and $M_2$ at their upper bounds (and neglecting additional terms in the parenthesis of eq.(\ref{mtbound1})) one obtains an upper bound for $\overline m_{\tilde t}$, namely $\overline m_{\tilde t}\simlt 1.7 \ {\rm~TeV}$. However, this is somehow too optimistic since it requires that all these HE parameters are simultaneously at their upper bounds, which is unlikely. A way to deal with this problem is to slightly modify the fine-tuning measure (\ref{BG}), in a (more restrictive) fashion, which counts all the contributions to the fine-tuning. Namely, instead using $\Delta\equiv {\rm Max}\ \left|\Delta_{\theta_i}\right|$, one defines $\Delta\equiv \{\sum_i \left|\Delta_{\theta_i}\right|^2\}^{1/2}$, which, as has been argued \cite{Casas:2005ev}, it is a more meaningful quantity. If the fine-tuning is dominated by one of the HE parameters (which is the usual case) both definitions are equivalent, but if there are several parameters contributing substantially to the fine-tuning, the second definition is more sensible (and restrictive). Then, it is easy to show that the maximum value of $\overline m_{\tilde t}^2$ subject to the condition $\Delta\leq \Delta_{\rm max} $, with $\Delta$ defined in this modified way, is
\be
\overline m_{\tilde t}^2=\left[2.972 ^2 (M_3^{\rm max})^4 +
0.339^2 (m_{Q_3}^{\rm max})^4+ 0.305^2 (m_{U_3}^{\rm max})^4+0.091^2(M_2^{\rm max})^4+\cdots
\right] ^{1/2}+ m_t^2\ .
\nonumber
\ee
Using just the dominant terms appearing explicitly above, we get (for $M_{\rm HE}=M_X$) $\overline m_{\tilde t}\leq 1320$ GeV.

From these results it is clear that for the unconstrained MSSM, with $M_{\rm HE}=M_X$, the naturalness bound on the gluino mass is much more important for LHC detection than the one on the stop masses. 
Next, we show the numerical values of the various naturalness bounds in a systematic way.

\section{Results for the unconstrained MSSM} \label{sect:unconstrained}

The unconstrained MSSM, where the soft-terms and $\mu-$term at the HE scale are taken as the independent parameters, has been already considered in the previous subsections as a guide to discuss the various naturalness bounds. However, we have so far restricted ourselves  to the case $M_{\rm LE}=1$~TeV, $M_{\rm HE}=M_X$. It is interesting to show the limits, both on the initial parameters and on the supersymmetric spectrum, for other choices of $M_{\rm HE}$. 
Following the procedure explained in subsections~\ref{sect:HEbounds} and \ref{sect:boundsspectra}, 
we have computed the fine-tuning constraints for three representative values of $M_{\rm HE}$, namely $M_{\rm HE}= 2\times10^{16}$~GeV, $10^{10}$~GeV and $10^4$~GeV, keeping $M_{\rm LE}=1$~TeV. Using the plots shown in the appendix the reader can evaluate the bounds for any other choice of $M_{\rm HE}$.

The absolute upper bounds on the most relevant HE parameters, obtained from eq.(\ref{IneqGen}), with the additional correction  (\ref{deltaM3}) for $M_3$, are shown in table \ref{tab:SoftMax}. Similarly, the corresponding bounds on supersymmetric masses at low energy, evaluated as in subsection~\ref{sect:boundsspectra}, are shown in table \ref{tab:PhysMax}. All the bounds have been obtained by setting $\Delta^{\rm max}=100$, they simply scale as $\sqrt{\Delta^{\rm max}/100}$.

\begin{table}[hbt]
 \centering
 {\small
\begin{tabular}{| l | c | c | c |}
\hline
~&$M_{\rm M_{\rm HE}}=2\times10^{16}$ &$M_{\rm M_{\rm HE}}=10^{10}$& $M_{\rm M_{\rm HE}}=10^{4}$ \\
 \hline
$M_3^{\rm max}(M_{\rm HE})$ & 660 & 1 162 &  5 376 \\
$M_2^{\rm max}(M_{\rm HE})$ & 1 646 & 1 750 &  3 500 \\
$M_1^{\rm max}(M_{\rm HE})$ & 8 002 & 6 100 &  11 048 \\
$A_t^{\rm max}(M_{\rm HE})$ & 2 504 & 2 227&  3 094\\
$m_{H_u}^{\rm max}(M_{\rm HE})$ & 1 038 & 1 046 &  913 \\
$m_{H_d}^{\rm max}(M_{\rm HE})$ & 6 945 & 14 472 & 9 784 \\
$\mu^{\rm max}(M_{\rm HE})$ & 624 & 640 &  630 \\
$m_{Q_3}^{\rm max}(M_{\rm HE})$ & 1 458 & 1 687 &  3 527 \\
$m_{U_3}^{\rm max}(M_{\rm HE})$ & 1 640 & 1 828 &  3 710 \\
$m_{D_3}^{\rm max}(M_{\rm HE})$ & 5 682 & 7 812 &  20 277 \\
$m_{Q_{1,2}}^{\rm max}(M_{\rm HE})$ & 5 601 & 7 693 &  19 288 \\
$m_{U_{1,2}}^{\rm max}(M_{\rm HE})$ & 3 818 & 5 254 &  13 975 \\
$m_{D_{1,2}}^{\rm max}(M_{\rm HE})$ & 5 613 & 7 722 &  19 764 \\
$m_{L_{1,2,3}}^{\rm max}(M_{\rm HE})$ & 5 557 & 7 664 &  20 278 \\
$m_{E_{1,2,3}}^{\rm max}(M_{\rm HE})$ & 5 524 & 7 607 &  20 278\\

\hline

\end{tabular}
}
\caption{Upper bounds on some of the initial (HE) soft terms and $\mu-$term for three different values of $M_{\rm HE}$, in the unconstrained MSSM scenario. All quantities are given in~GeV units.}
\label{tab:SoftMax}
\end{table}

\begin{table}[hbt]
 \centering
 {\small
\begin{tabular}{| l | c | c | c |}
\hline
~&$M_{\rm HE}=2\times10^{16}$ &$M_{\rm HE}=10^{10}$& $M_{\rm HE}=10^{4}$ \\
 \hline
$M_{\tilde g}^{\rm max}$ &1 440 & 1 890 &  5 860 \\
$M_{\tilde W}^{\rm max}$ & 1 303 &1 550 &  3 435\\
$M_{\tilde B}^{\rm max}$ & 3 368 & 4 237 &  10 565 \\
$M_{\tilde H}^{\rm max}$ & 627 &  627 &   627 \\
$\overline m_{\tilde t}^{\rm max}$ & 1 320 & 1 590 &  3 190 \\
$ m_{H^0}^{\rm max}$ & 7 252 & 14 510 &  9 900 \\
\hline
\end{tabular}
}
\caption{Upper bounds on some of the physical masses for three different values of $M_{\rm HE}$, in the unconstrained MSSM scenario. All quantities are given in~GeV units. 
}
\label{tab:PhysMax}
\end{table}

\noindent
From the previous tables we can notice some generic facts.  

\begin{itemize} 

\item
Taking into account the present and future LHC limits, the upper bound on the gluino mass is typically the most stringent one, being at the reach of the LHC (for $\Delta^{\rm max}=100$), unless the high-energy scale is rather low. On the other hand, the gluino bound is  the most sensitive to the value of $M_{\rm HE}$, since it is a two-loop effect. For $M_{\rm HE}\simeq10^7$~GeV, it is as already beyond the future LHC limit ($\sim 2.5$~TeV, see e.g.  \cite{LHC-talk})  and it increases rapidly as $M_{\rm HE}$ approaches the electroweak scale. 

\item
The upper bound on the wino mass, $M_{\tilde W}$, is similar to the gluino one. Note here that (unless $M_{\rm HE}$ is quite small) the weight of $M_2^2$ in the value of $m_{H_u}^2(M_{\rm LE})$ is certainly smaller than that of $M_3^2$; but this effect is compensated, when computing the physical masses, by the large increase of $M_3$ when running from $M_{\rm HE}$ to $M_{\rm LE}$ (see Figs.~1, 4). On the other hand, the bound on $M_{\tilde W}$ is much less restrictive than the one on $M_{\tilde g}$, given the LHC  discovery potential. The upper bound on the bino, as expected, is quite mild and always beyond the reach of the next LHC run. This is just a consequence of the little impact that $M_1$ has on $m^2_{H_u}(LE)$.

\item
Concerning electroweakinos, the most relevant upper bounds are those on Higgsinos, $M_{\tilde H}$. Not only they are the strongest ones (hopefully at the reach of the LHC for $\Delta^{\rm max}=100$), but also, by far, the most stable of all bounds. This is a consequence of the fact that $\mu$ runs proportional to itself, so their fine-tuning parameter is insensitive to the HE scale, see eq.(\ref{BGmu}). Apart from that, the running of $\mu$ is very little. It is worth-mentioning that for $\Delta^{\rm max}=100$ the upper bound on $M_{\tilde H}$ is not far from $M_{\tilde H}\simeq 1$~TeV, which is the value required if dark matter is made of Higgsinos \cite{ArkaniHamed:2006mb, Cahill-Rowley:2014boa}.

\item
The upper bounds on stops are not as stringent as the gluino one unless $M_{\rm HE}$ is pretty close to the electroweak scale, in which case none of them is relevant. In general, it is not justified to say that Natural SUSY prefers light stops, close to the LHC limits. Actually, for $\Delta^{\rm max}=100$ the upper bounds on stops are beyond the LHC reach \cite{LHC-talk}. Taking lighter stops does not really improve the fine-tuning since there are other contributions to it which are dominant, in particular the gluino one.

\item
Given the present LHC limits, the contribution of the gluino to the $m_{H_u}^2$ is bigger than that of stops, then it is not useful to have light stops. This conclusion is reinforced when other aspects are considered, see subsection~\ref{sect:125} below. Unless $M_{\rm HE}$ is very small, the gluino mass sets the level of EW fine-tuning of the unconstrained MSSM, which is  ${\cal O}(1\%)$.

If $M_{\rm LE}$ becomes close to the electroweak scale, the supersymmetric fine-tuning becomes much less severe. This fact is strengthened by the fact that additional soft dimension-4 Higgs operators may start to become relevant, increasing the tree-level Higgs mass and thus decreasing further the fine-tuning. These aspects were noted in 
ref.~\cite{Brignole:2003cm,Casas:2003jx, Dine:2007xi}.

\item 
Concerning the squarks of the first two generations and all the generations of sleptons, their bounds are, as expected, far beyond the reach of the LHC; the reason being that their contribution to $m_{H_u}^2$ is very small.

\item
Lastly, we can see the large upper bounds on $m_{H_d}^2$. When $M_{\rm LE}$ is very large, its contribution to the fine-tuning is very small. However, for low values of $M_{\rm LE}$, the term proportional to $m^2_{H_d}(\tan^2\beta-1)^{-1}$ (neglected for simplicity in expr.(\ref{minh})) actually becomes the dominant one, causing a larger impact of $m_{H_d}^2$ on the EW fine-tuning and, as consequence, decreasing the respective upper bound. This can be seen from table (\ref{tab:SoftMax}), where the bound on $m_{H_d}$ gets lower for $M_{\rm LE}=10^4$~GeV. Being the largest bounds as compared to the ones of $\mu$ and $m^2_{H_u}$, the term $m^2_{H_d}$ dominates the bounds on the masses of the heavy Higgses (see table \ref{tab:PhysMax}).

\end{itemize}

\section{Impact of other potential fine-tunings of the MSSM} \label{sect:impactextraFT}

\subsection{Fine-tuning to get $m_h^{\rm exp}\simeq 125$~GeV} \label{sect:125}

\begin{figure}[ht]
\centering
 \includegraphics[width=0.75\linewidth]{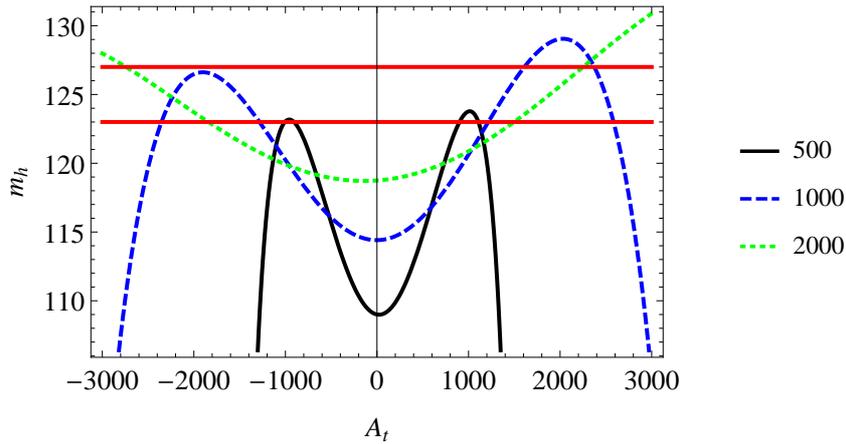}
 \caption{The Higgs boson mass, $m_h$, as a function of the third generation squark masses, $m_{Q_3} = m_{U_3}$. The black-solid line is for $m_{Q_3} = 500$~GeV, blue-dashed for $m_{Q_3} = 1000$~GeV, and green-dotted for $m_{Q_3} = 2000$~GeV. The red horizontal lines denote $m_h = 125 \pm 2$~GeV band. The Higgs boson mass has been calculated using
 \texttt{FeynHiggs~2.10.1}~\cite{Hahn:2013ria,  Frank:2006yh, 
 Degrassi:2002fi,  Heinemeyer:1998np, Heinemeyer:1998yj}. \label{fig:higgs_mass}} 
\end{figure}

From the results of the previous section it is clear that, concerning naturalness, little is gained by going to light stops, say $< 800$~GeV. Actually, such light stops could entail, as already mentioned in section~\ref{sect:extraFT}, an additional fine-tuning since the condition $m_h^{\rm exp}\simeq 125$~GeV may require the threshold contribution to the Higgs mass to be maximal with high accuracy.
The relevant equation is
\begin{eqnarray}
\label{mh}
m_h^2=(m_h^2)_{\rm tree-level} \ +\ \delta_{\rm rad}m_h^2\ +\  \delta_{\rm thr}m_{h}^{2}\ ,
 \end{eqnarray}
where $\delta_{\rm rad}m_h^2$ ($\delta_{\rm thr}m_{h}^{2}$) is the radiative (threshold) contribution to $m_h^2$, approximately given by the $X_t-$independent (dependent) part of eq.(\ref{eq:der2}). We recall that for moderately large $\tan\beta$ one can approximate $X_{t}=A_{t}(M_{\rm LE})-\mu \cot\beta\simeq A_t(M_{\rm LE})$. Figure~\ref{fig:higgs_mass} shows the dependence of $m_h$ vs $A_t(M_{\rm LE})$ for different values of the (LE) soft stop-masses, taken as degenerate for simplicity, $m_{Q_3} = m_{U_3} = 500, 1000, 2000$~GeV. If the stops are light, $\sim 500$~GeV, the
correct value of the Higgs boson mass, $m_h = 125 \pm 2$~GeV (the uncertainty is mainly due to the theoretical calculation), requires $A_t(M_{\rm LE})$ to be precisely fine-tuned\footnote{Note that in this case the ``standard criterion" to evaluate the fine-tuning, i.e. $\Delta=\partial \log m_h/\partial\log A_t$ is not applicable (indeed, one would conclude from it that there is no fine-tuning at all), since $A_t$ is close to an stationary point, see footnote 5.} at $\pm 1000$~GeV. On the other hand, if the stop masses are $\sim 1000$ or $2000$~GeV, a broad range of values is allowed, $A_t(M_{\rm LE}) = \pm (2000 \pm 1000)$~GeV, which entails no fine-tuning.

We emphasize that this potential fine-tuning is independent of the one required to obtain the correct electroweak scale, which has been analyzed in the previous section. Therefore, if both fine-tunings are present we should combine them, i.e.\ multiply the two small probabilities of getting both the correct electroweak scale and the correct Higgs mass. This requires to quantify the fine-tuning associated to the Higgs mass in a fashion which has similar statistical meaning as the measure used for the electroweak fine-tuning. Taking into account the discussion of subsection~\ref{sect:FTmeasure}, we adopt here a fine-tuning  measure that is also consistent with an interpretation in terms of $p$-value. 
In particular, if the stops are light, the fine-tuning is well reflected by the $p-$value of getting $m_h$ as large as $m_h^{\rm exp}$ or larger
\begin{eqnarray}
\label{pvaluemh}
p-{\rm value} = \int_{m_h\geq m_h^{\rm exp}}\ d m_h\ {\cal P} (m_h) \ .
\end{eqnarray}
Here  $ {\cal P} (m_h)$ is the probability of a Higgs mass value, given by
\begin{eqnarray}
\label{Pmh}
{\cal P} (m_h)=\left| \frac{dX_t}{dm_h}\right| {\cal P}(X_t(m_h)) \ ,
 \end{eqnarray}
where ${\cal P} (X_t)$ is the probability distribution of $X_t-$values.\footnote{For many values of $m_h$, there are four $X_t$ solutions, so ${\cal P} (m_h)$ is the sum of four terms, corresponding to those solutions.} The final step is to assume a shape for ${\cal P} (X_t)$. Note here that $X_t\simeq A_t(M_{\rm LE})$ is a low-energy quantity, so it is not much sense to adopt a prior for it. Strictly speaking, the prior should be assumed for the initial, high-energy parameters that determine the value of $A_t(M_{\rm LE})$, (i.e.\ $A_t, M_3, M_2$), in a similar fashion as the one followed to establish the electroweak fine-tuning in the previous sections. Nevertheless, it is clear from figure~\ref{fig:higgs_mass} that, roughly speaking, for $\overline m_{\tilde t} \gtrsim 1000$~GeV and any sensible theoretical scenario for the soft terms, the $p-$value will be $\sim 20\%$ or larger, which means that there is not really a fine-tuning associated to $m_h\simeq 125$~GeV. Living in this range, the only important 
fine-tuning is 
the one associated to the electroweak scale. On the other hand, if stops are very light, both fine-tunings should be simultaneously considered.  Then, one should multiply the $\Delta_{\rm electroweak}$ parameter by the inverse of the above $p$-value, which necessarily leads to a per-mil (or even more severe) global fine-tuning. So, interestingly, if  the average stop mass is light, say $\simlt 800$~GeV, the situation is typically more fine-tuned than for heavier stops, 
 $\sim {\cal O}(1\ {\rm~TeV})$.

\subsection{The Higgs mass and the parameter space selected by naturalness} \label{sect:Hmass}

 \vspace{0.2cm}
On the other hand, even if there is no fine-tuning to get the experimental Higgs mass, the requirement $m_h=m_h^{\rm exp}$ implies a balance between $\delta_{\rm rad}m_h^2$ and $\delta_{\rm thr}m_{h}^{2}$ in eq.(\ref{mh}), which in turn implies a correlation between the initial parameters, especially $M_3$ (the main responsible for the size of the stop masses) and $A_t$. This correlation has non-trivial consequences for the electroweak fine-tuning.

To see this, consider $\Delta_{M_3}$, which is usually the most significant fine-tuning parameter. As discussed in subsection~\ref{sect:HEbounds}, $\Delta_{M_3}$ is a function, not only of $M_3$, but also of $M_2$ and $A_t$. E.g. for $M_{\rm HE}=M_X, M_{\rm LE}= 1$~TeV, $\Delta_{M_3}$ is given by eq.(\ref{deltaBGM3}), where one can note that it will get partially suppressed as long as $M_3$ and $A_t$ are of the same sign. Therefore, fixing $M_3 > 0$ one would expect the lowest electroweak fine-tuning for $A_t > 0$. On the other side, it is evident from table~\ref{tab:trilinears} that the RG running pushes such $A_t$ towards rather low and possibly negative values. However, low values of $A_t$ at LE are in conflict with the measured Higgs boson mass, as can be seen in figure~\ref{fig:higgs_mass}. This will result in a tension between low fine tuning of the electroweak scale and the Higgs mass.    

\begin{figure}[ht] 
\centering 
\includegraphics[width=0.45\linewidth]{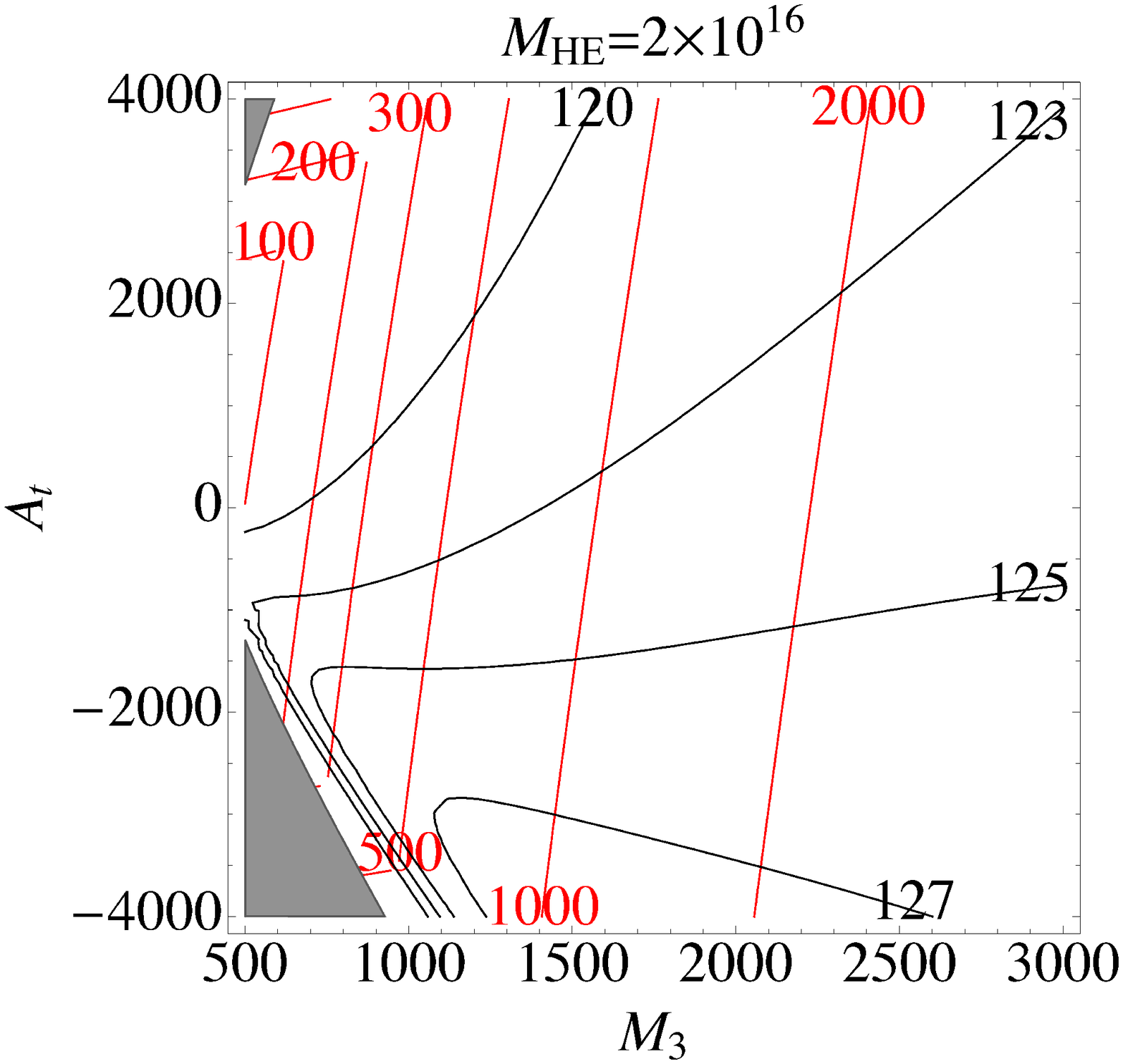}
\includegraphics[width=0.45\linewidth]{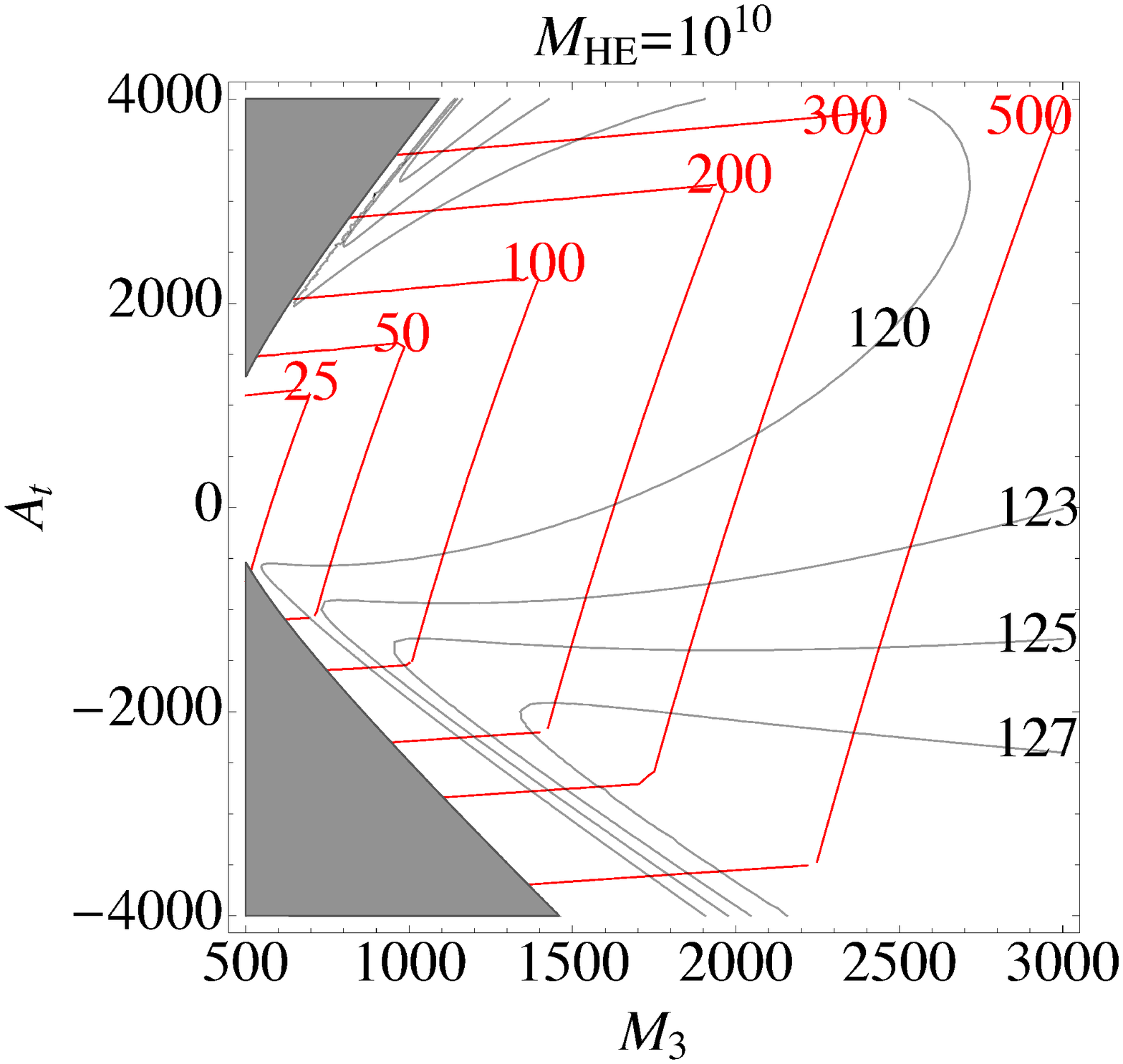}\\
\includegraphics[width=0.45\linewidth]{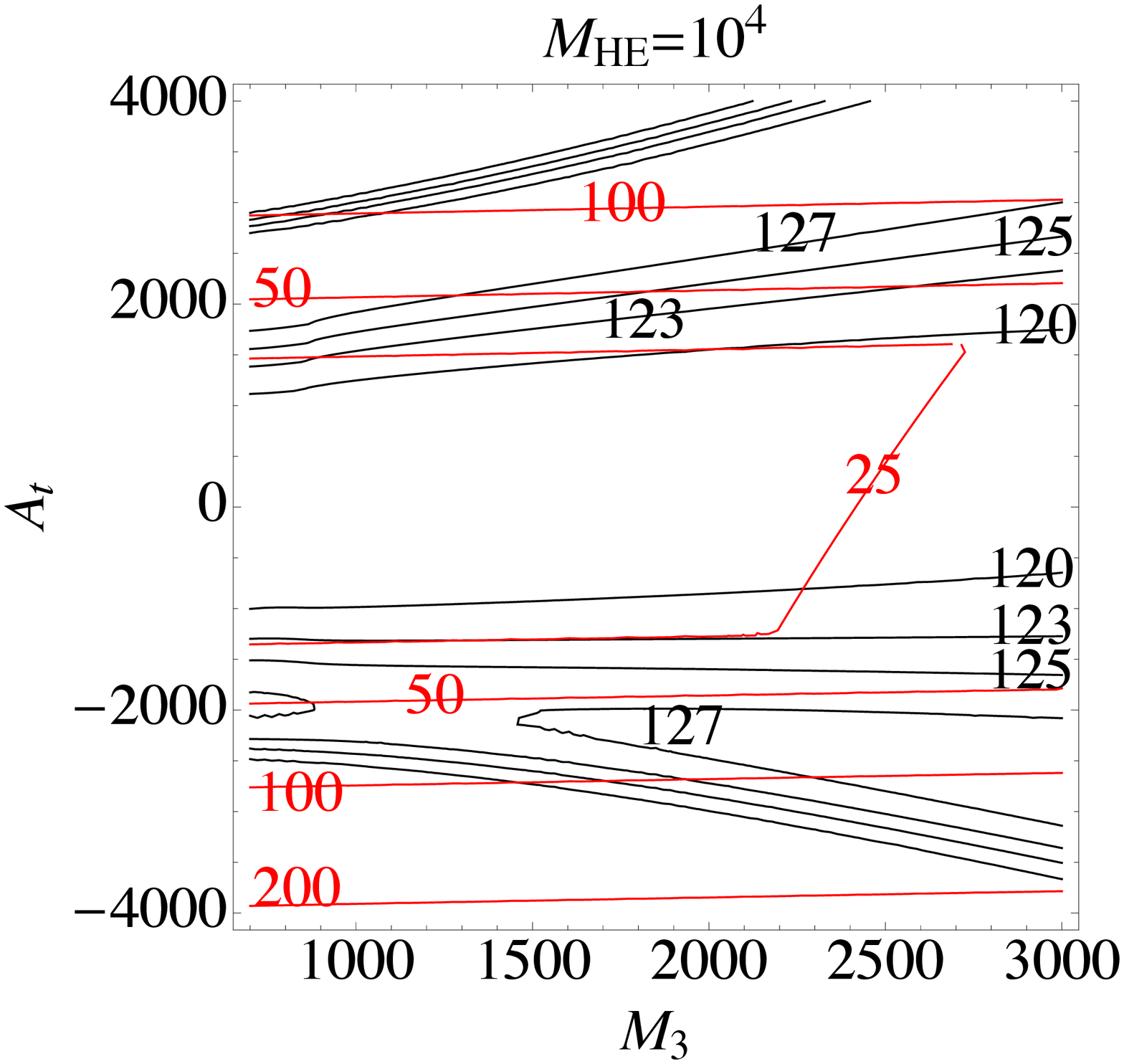}
\caption{Contours of constant Higgs boson mass (black, contours for $m_h = 120,123,125,127$~GeV) and fine-tuning (red), eq.~\eqref{BGTheta}, in the $M_3$--$A_t$ plane. We have chosen $M_2=M_3$ and $m_{Q_3}^2=m_{U_3}^2=0$ at HE. From left to right to bottom, $M_\mathrm{HE} = 2\cdot 10^{16}, 10^{10}, 10^4$~GeV. The unphysical region with tachyonic stops is shaded in gray. 
\label{fig:mh_finetuning}} 
\end{figure}

This situation is depicted in figure~\ref{fig:mh_finetuning} where we show the contours of constant Higgs boson mass (black) and fine tuning (red), together in the (high-energy) $M_3$--$A_t$ plane, for different choices of $M_\mathrm{HE}$. For simplicity we have chosen $M_2=M_3$ and $m_{Q_3}^2=m_{U_3}^2=0$ at HE. Note here that, unless the HE stop masses are very large, their LE values are essentially determined by $M_3$ (unless $M_\mathrm{HE}$ is small), so the results of the figures are quite general.
The fine-tuning shown corresponds to the largest  $\Delta$ among the parameters. Usually it is given by $\Delta_{M_3}$, especially when there is a significant amount of running, although for large $|A_t|$ it may be given by $\Delta_{A_t}$ (then the red lines get horizontal in the plots).
As expected from the above discussion, when the fine-tuning is dominated by $\Delta_{M_3}$, it  tends to be lower for $A_t > 0$;
however, $m_h \sim 125$~GeV prefers  $A_t < 0$. For $M_\mathrm{HE} = 2\cdot 10^{16}$~GeV (upper-left panel of figure~\ref{fig:mh_finetuning}), the Higgs boson mass requires $A_t\sim -2000$~GeV, resulting in a large fine tuning, $\Delta \sim 250$. Moreover, $M_3$ is required to be larger than $\sim 750$~GeV which implies that the gluino mass should be (at least) slightly above current exclusion limits. Of course larger values of $M_3$ result in a more severe fine-tuning, as is clear from the figure. The tension between different low energy requirements is clearly visible in the upper-right panel, $M_\mathrm{HE} = 10^{10}$~GeV, where the correct Higgs mass is obtained for $A_t \sim -1500$~GeV with $\Delta\sim 100$ or even smaller,  which corresponds to  $M_3\sim 900$ GeV and, again, a physical gluino mass  just above the current exclusion limits. Once more, higher values of the gluino mass imply higher fine-tuning, but the increase is not as dramatic as for $M_\mathrm{HE} = 2\cdot 10^{16}$~GeV. On the other hand, for positive $A_t$ a much higher value is required, $A_t \sim 3000$~GeV, 
which results in a significant increase in fine tuning due to $A_t$, namely $\Delta_{A_t} \sim 300$. 
Only for a very low choice of the high-energy scale, $M_\mathrm{HE} = 10^4$~GeV, the positive $A_t$ is preferred. In this case the fine-tuning gets substantially smaller,  $\Delta \lesssim 50$.
The result is rather independent of $M_3$ which only enters at 2-loops in the Higgs mass and has a very limited impact on other SUSY parameters due to RGE running.

We can therefore conclude that,  unless the scale of SUSY breaking transmission is quite low, the least fine-tuned scenarios (i.e. the most ``natural'' ones) generically demand negative $A_t$, a requirement driven by the measured Higgs mass. The corresponding fine-tuning is ${\cal O}(100)$, with gluinos only slightly heavier than the current limits, promising interesting discovery prospects at the second run of the LHC with increased center-of-mass energy.  

\subsection{Fine-tuning to get large $\tan\beta$}\label{tanbeta}

As pointed out in subsection~2.3, a large value of $\tan\beta$ generically requires a small value of $B\mu$ at low energy, which requires a cancellation between the initial value and the radiative contribution from the RG-running. Here, we quantify this fine-tuning and discuss its consequences.

From eq.(\ref{B}), we can write, for $\tan\beta\gg 1$,
\bea
\label{tB}
\tan\beta\simeq  \frac{m_{H_d}^2 + m_{H_u}^2+2\mu^2}{B\mu}=\frac{m_A^2}{B\mu}\ ,
\eea
where $m_A$ is the mass of the pseudoscalar Higgs and all the quantities are understood at the low-energy (LE) scale. As discussed in subsection~\ref{sect:extraFT}, the fine-tuning to get large $\tan\beta$ can be reasonable quantified using the standard criterion. Namely, for any initial parameter of the theory, $\theta$, we define the associated fine-tuning, $\Delta_\theta^{(\tan\beta)}$
\bea
\label{DeltatB1}
\Delta_\theta^{(\tan\beta)}=\frac{\theta}{\tan\beta}\frac{d \tan\beta}{d \theta}=\frac{\theta}{m_A^2}\left[\frac{d m_A^2}{d\theta}-\tan\beta\ \frac{d (B\mu)}{d \theta}\right]\ ,
\eea
where we have used eq.(\ref{tB}). For large $\tan\beta$, $\Delta_\theta^{(\tan\beta)}$  is normally dominated by the second term within brackets in  (\ref{DeltatB1})
\bea
\label{DeltatB}
\left|\Delta_\theta^{(\tan\beta)}\right|\simeq  \tan\beta\left|\frac{\theta}{m_A^2}\frac{d (B\mu)}{d \theta}\right|\ .
\eea
The next step is to express the LE value of $B\mu$ in terms of the initial (HE) parameters. E.g. assuming $M_{\rm HE}=M_X$, $M_{\rm LE}=1$~TeV, from table~\ref{tab:mu-Bmu}
\bea
\label{Bmulow}
B\mu (LE)\simeq B\mu + 0.46 M_3\mu -0.35M_2\mu-0.34 A_t \mu-0.03M_1\mu+\cdots ,
\eea
where the quantities on the r.h.s. are at the HE scale.  Then, the corresponding fine-tuning  $\Delta$s for the relevant parameters\footnote{Note that $\Delta_\mu^{(\tan\beta)}\simeq 1$.}, $B, M_3,M_2,A_t$, read
\bea
\label{DeltatBs}
\left|\Delta_{\{B, M_3,M_2,A_t\}}^{(\tan\beta)}\right|\simeq  \tan\beta\left| \frac{\mu}{m_A^2}\{B, \ 0.46M_3,\ 0.35M_2, \ 0.34 A_t  
\}\right| \ ,
\eea
where we recall that r.h.s. parameters at the HE-scale. Going to particular models, one clearly expects some of the $\{\mu B, \ \mu M_3,\ \mu M_2, \ \mu A_t 
\}$ quantities to be of the order of $m_A^2$. Indeed, the HE value of $B$ could be zero, but $M_3,M_2$ cannot.
This means that a certain fine-tuning, $\Delta^{(\tan\beta)}\simgt 5-10$ occurs if $\tan\beta\simgt 15-30$. Since this fine-tuning has a different nature from the electroweak one (discussed in detail in the previous sections), and given the probabilistic meaning of the fine-tuning parameters, this implies that the two $\Delta$s have to be multiplied, $\Delta=\Delta^{({\rm EW})}\Delta_\theta^{(\tan\beta)}$, which generically results in an exaggerated fine-tuning ($> 500-1000$). Notice that these conclusions are alleviated if the HE scale is smaller, since the numerical coefficients in (\ref{Bmulow}) decrease.
On the other hand, for $\Delta^{(\tan\beta)}\simlt 5$ there is no really fine-cancellation to get the value of $\tan\beta$ and one can ignore the $\Delta^{(\tan\beta)}$ fine-tuning factor.

The conclusion is that very large $\tan\beta$, say $\tan\beta\simgt 15-30$, implies a high fine-tuning price, unless the special characteristics of the model lead to a small r.h.s. in (\ref{DeltatB}), e.g. if $m_A^2$ is abnormally large. 

Let us conclude this section pointing out that for $\tan\beta\simgt 30$ the impact of the bottom and tau Yukawa couplings in the RGEs become non-negligible, so the previous numerical values would be modified, but the general conclusion would be the same.

 \section{Summary and Conclusions: the most robust predictions of a Natural SUSY scenario} \label{sect:Summary} 

The idea of ``Natural SUSY", understood as a supersymmetric (MSSM) scenario where the fine-tuning is as mild as possible, is a reasonable guide to explore supersymmetric phenomenology,  since, as usually argued, the main phenomenological virtue of SUSY is precisely to avoid the huge fine-tuning associated to the hierarchy problem. Much work has been done in the literature to quantify the fine-tuning of a generic MSSM and to extract the features of Natural SUSY. However, these analyses often ignore relevant aspects, such as the ``mixing" of the fine-tuning conditions or the presence of other potential fine-tunings.

In this paper, we have addressed the supersymmetric fine-tuning in a comprehensive way, including the discussion of the measure of the fine-tuning and its probabilistic meaning, the mixing of  the fine-tuning conditions, the method to extract fine-tuning bounds on the initial parameters and the low energy supersymmetric spectrum, as well as the role played by extra potential fine-tunings. We have given tables and plots that allow to easily evaluate the fine-tuning and the corresponding naturalness bounds for any theoretical model defined at any high-energy (HE) scale. Finally, we have analyzed in detail the complete fine-tuning bounds for the unconstrained MSSM, defined at any HE scale, including the impact that the experimental Higgs mass imposes on  the soft terms.

From the results of the previous sections, we summarize below the most important implications of fine-tuning in the MSSM; or, in other words, the characteristics of a Natural-SUSY scenario.

\begin{enumerate}

\item 
For the evaluation of the fine-tuning it is crucial to define: i) the initial (independent) parameters of the theoretical setup, ii) the high-energy (HE) scale at which they are defined and iii) the criterion to quantify the fine-tuning. 

We have seen that the `standard' fine-tuning criterion (\ref{BG}) normally has a sound statistical meaning, though one should be careful about the implicit assumptions of the prior for the initial parameters hold (if not, the standard criterion has to be consistently modified). Besides, we have provided tables and plots (see the appendix) that allow to  straightforwardly  evaluate the fine-tuning for any theoretical setup at any HE-scale.

\item
Concerning the electroweak fine-tuning of the MSSM (i.e. the one required to get the correct electroweak scale), the most robust result is by far that Higgsinos should be rather light, certainly below 700~GeV for $\Delta<100$, i.e. to avoid a fine-tuning stronger than 1\% (all the bounds on masses scale as $\sqrt{\Delta^{\rm max}}$). This result is enormously stable against changes in the HE scale since the $\mu-$parameter runs proportional to itself (besides running very little from HE to LE). The only way it could be substantially relaxed would be that the $\mu-$parameter were theoretically related to the soft masses in such a way that there occurred a cancellation at LE between $\mu^2$ and $m_{H_u}^2$ (see eq.(\ref{minh})). This is difficult to conceive and, certainly, it is not realized in the known theoretical SUSY frameworks. Incidentally, this upper bound is not far from $M_{\tilde H}\simeq 1$~TeV, which is the value required if dark matter is made of Higgsinos.

\item
The most stringent naturalness upper bound, from the phenomenological point of view, is the one on the gluino mass. If $M_{\rm HE} \simeq M_X$ one gets $M_{\tilde g}\simlt 1.5$~TeV for $\Delta^{\rm max}=100$, i.e. just around the corner at the LHC. In other words, the gluino mass typically sets the level of the electroweak fine-tuning of the MSSM, which at present is  ${\cal O}(1\%)$.

However, this limit is not as robust as the one on Higgsinos. First, it presents a strong dependence on the HE-scale (due to the two-loop dependence of the electroweak scale on the gluino mass). Actually, for $M_{\rm HE}\simlt10^7$~GeV and $\Delta^{\rm max}=100$ the upper bound on $M_{\tilde g}$ (about 2.7~TeV) goes beyond the present LHC reach. In addition, it could be relaxed if the initial soft parameters (e.g. the gaugino masses) are theoretically related in a favorable way.

\item
The upper limit on the wino mass, $M_{\tilde W}$, is slightly smaller than the gluino one, but less relevant for LHC phenomenology. It also has a similar degree of robustness, though it is less dependent on $M_{\rm HE}$. The upper bound on the bino mass, $M_{\tilde B}$ is weaker and beyond the LHC reach.

\item 
A remarkable conclusion is that light stop masses are {\em not} really a generic requirement of Natural SUSY. Actually, stops could be well beyond the LHC limits without driving the electroweak fine-tuning of the MSSM beyond 1\%. Even more, in some scenarios, like universal scalar masses with $M_{\rm HE}=M_X$, stops above 1.5~TeV are consistent with a quite mild fine-tuning of $\sim$ 10\%. Hence, the upper bounds on stops are neither stringent nor stable under changes of the theoretical scenario.

In contrast, as mentioned above, the gluino mass is required to be light with much more generality, although its impact on the fine-tuning depends crucially on the size of $M_{\rm HE}$ (it is maximum for $M_{\rm HE}=M_X$). Consequently, the electroweak scale is typically fine-tuned at 1\% in most cases, and having light stops does not help, since the electroweak fine-tuning stems from a single cancellation between terms, essentially between the ones proportional to $M_3^2$ and $\mu^2$ in eq.(\ref{minh}).

\item 
In addition to the conventional fine-tuning to get the correct electroweak scale, there are two potential extra fine-tunings, namely the tuning of the threshold correction to get $m_h=m_h^{\rm exp}$ when stops are too light, and the tuning of $B\mu$ (at low energy) to get a large $\tan\beta$. It is convenient to avoid these additional fine-tunings, otherwise they have to be combined with (i.e. multiplied by) the electroweak fine-tuning, normally resulting in a gigantic global fine-tuning. Typically, this requires a not-too-light average stop mass, i.e. $\overline m_{\tilde t}\simgt 800$~GeV; and not-too-large $\tan\beta$, i.e. $\tan\beta\simlt 15-30$. The precise conditions to avoid these tunings are discussed in sect.~\ref{sect:impactextraFT}. Note that a small average stop mass is disfavored, but the mass of the lightest stop could be light or very light.

\item
Unless the high-energy scale  is quite low, the less fine-tuned scenarios generically demand negative $A_t$, a requirement driven by the measured Higgs mass. The corresponding fine-tuning is ${\cal O}(100)$, with gluinos only slightly heavier than the current limits, which offers interesting prospects for the second run of the LHC.

\item 
Lastly, the fine-tuning bounds on all the sleptons, the first two generations of squarks and the heavy Higgs states, are, as expected, far beyond the reach of LHC. This is a consequence of the little effect these parameters have on the value of $m_{H_u}^2$ at low energy. 
\end{enumerate}

\section*{Acknowledgements}

The authors want to thank L. Calibbi, D. Cerdeno, S. Heinemeyer and L. Ibanez for very useful discussions. 
This work has been partially supported by the MICINN, Spain, under contract FPA2010-17747, FPA2013-44773-P, Consolider-Ingenio CPAN CSD2007-00042, as well as MULTIDARK CSD2009-00064. 
We also thank the Spanish MINECO {\em Centro de excelencia Severo Ochoa Program} under
grant SEV-2012-0249. 
B.Z. is supported by the IISN, an ULB-ARC grant and by the Belgian Federal Science Policy
through the Interuniversity Attraction Pole P7/37,  S.R.  by Campus of Excellence UAM+CSIC and  K.R.   by Spanish Research Council (CSIC) within the JAE-Doc program.

\newpage

\appendix
\section{Low-energy running coefficients at 2 loops}

We compile in this appendix the coefficients of the functional forms that exactly fit the low-energy (LE) parameters in terms of the high-energy (HE) ones.  Namely, for dimension-two parameters, say ${\cal M}^2$
\begin{eqnarray}
\label{m2_gen_fit}
{\cal M}^2(LE)&=&
c_{M_3^2}M_3^2 +c_{M_2^2}M_2^2 +c_{M_1^2}M_1^2 + c_{A_t^2}A_t^2+c_{A_tM_3}A_tM_3 +c_{M_3M_2}M_3M_2+ \cdots 
\nonumber\\
&&\cdots +c_{m_{H_u}^2}m_{H_u}^2 +c_{m_{Q_3}^2} m_{Q_3}^2 +c_{m_{U_3}^2}m_{U_3}^2+\cdots ,
\end{eqnarray}
where the r.h.s. parameters are understood at the HE scale. Similarly, for dimension-one parameters, say ${\cal M}$, we have
\begin{eqnarray}
\label{m_gen_fit}
{\cal M}(LE)=
c_{M_3}M_3 +c_{M_2}M_2 +c_{M_1}M_1 + c_{A_t}A_t+ \cdots 
\end{eqnarray}

In tables \ref{tab:mhiggssq}--\ref{tab:mu-Bmu}  we list the values of  the above $c-$coefficients for each LE soft term and for the LE $\mu-$parameter. These values correspond to the choice $M_{\rm HE}=M_X$, $M_{\rm LE}=1$~TeV and $\tan\beta=10$. 

The dependence on $\tan\beta$ is very small provided $5\simlt \tan\beta\simlt 30$. If $\tan\beta\simlt 5$ the top Yukawa coupling becomes larger, affecting the entire set of RGEs. Likewise, for larger values of $\tan\beta\simgt 30$ the effect of the bottom and tau Yukawa couplings start to be non-negligible. Notice however $\tan\beta\simlt 5$ implies extremely heavy stops, so that the radiative correction to the Higgs mass is large enough to reproduce $m_h\simeq 125$~GeV. This amounts to an enormous fine-tuning. Analogously, for $\tan\beta\simgt 30$ the tuning required to get large $\tan\beta$ usually raises the global fine-tuning up to unreasonable levels, see sect.~\ref{sect:extraFT}.

The dependence of the $c-$coefficients on $M_{\rm LE}$ is logarithmic and can be well approximated by
\begin{eqnarray}
\label{scaledep_app}
c_i(M_{\rm LE}) \simeq c_i(1\ {\rm~TeV}) + b_i\ln\frac{M_{\rm LE}}{1\ {\rm~TeV}} \ .
\end{eqnarray}
The value of the $b_i$ coefficients is also given in Tables \ref{tab:mhiggssq}--\ref{tab:mu-Bmu} . 

\begin{figure}[h!]
\centering 
\includegraphics[width=0.8\linewidth]{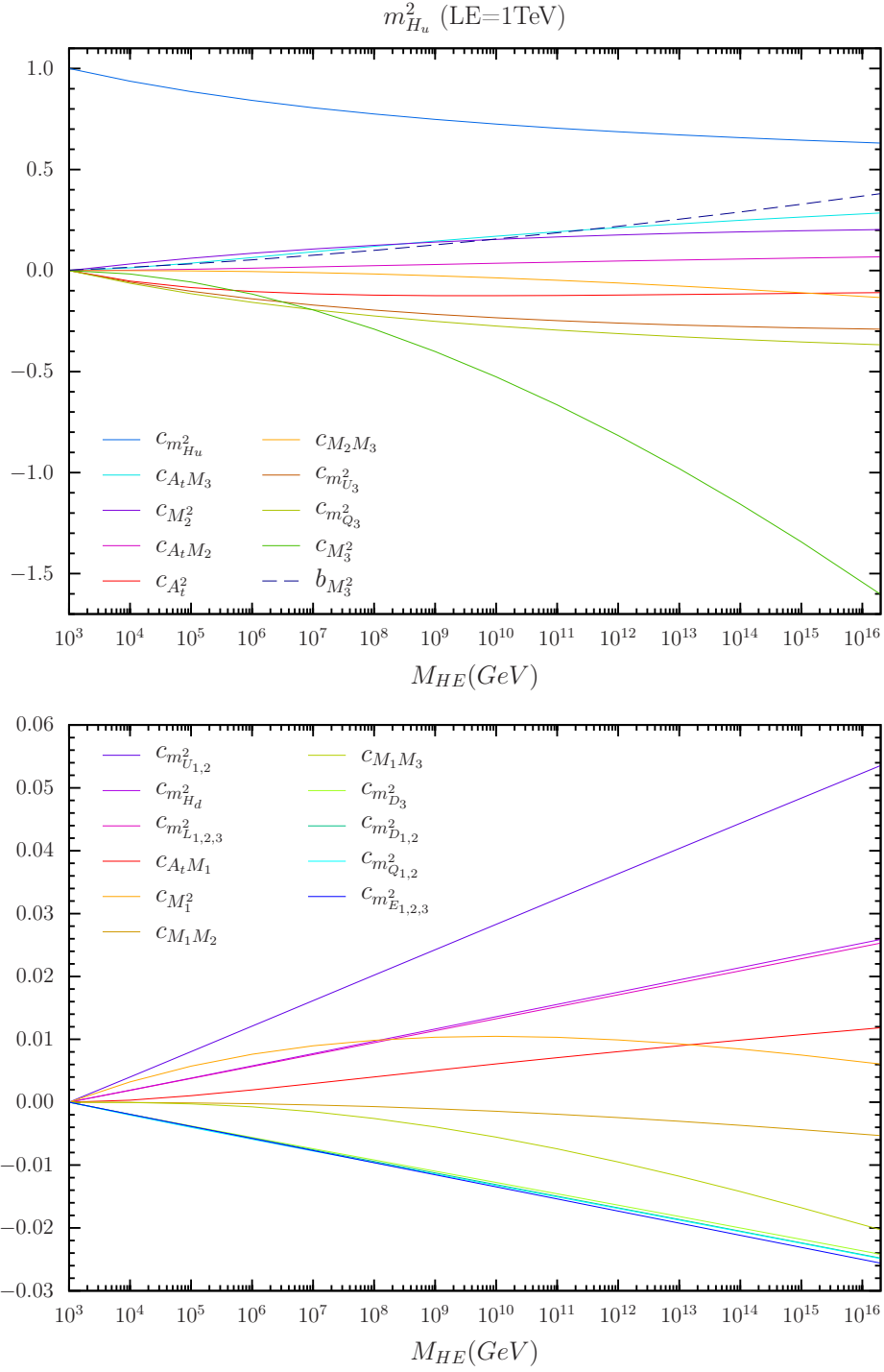}
\caption{$m_{H_u}^2(M_{\rm LE})$ coefficients dependence on the HE scale, for $M_{\rm LE}=1\TeV$ and $\tan \beta = 10$ . For further details,  see eqs.~(\ref{m2_gen_fit}--\ref{scaledep_app}).} 
\label{fig:mHusq}
\end{figure}

\begin{figure}[h!]
\centering 
\includegraphics[width=0.8\linewidth]{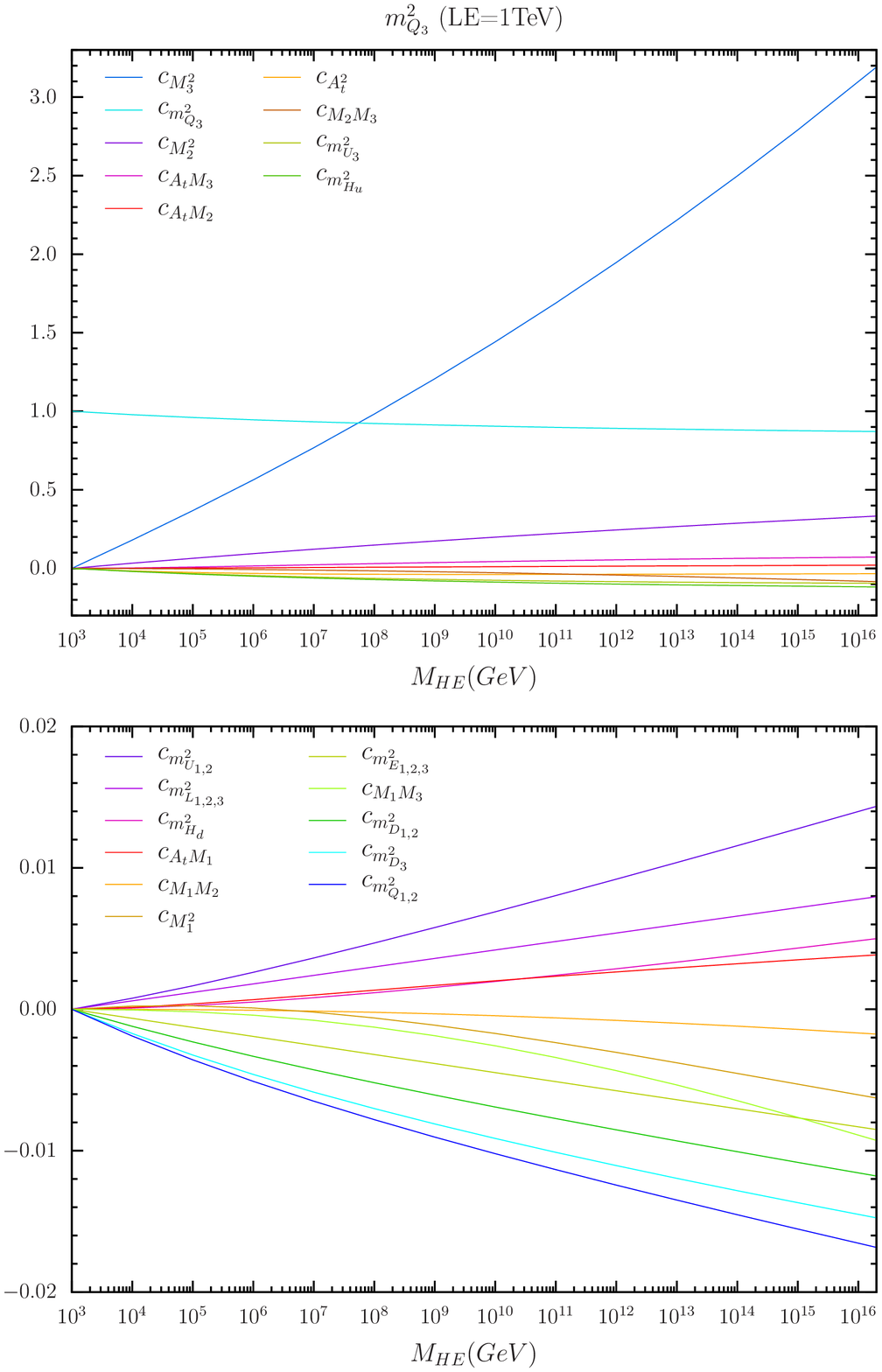}
\caption{As Figure 3, for $m_{Q_3}^2(M_{\rm LE})$.} 
\label{fig:mQ3sq}
\end{figure}

\begin{figure}[h!]
\centering 
\includegraphics[width=0.8\linewidth]{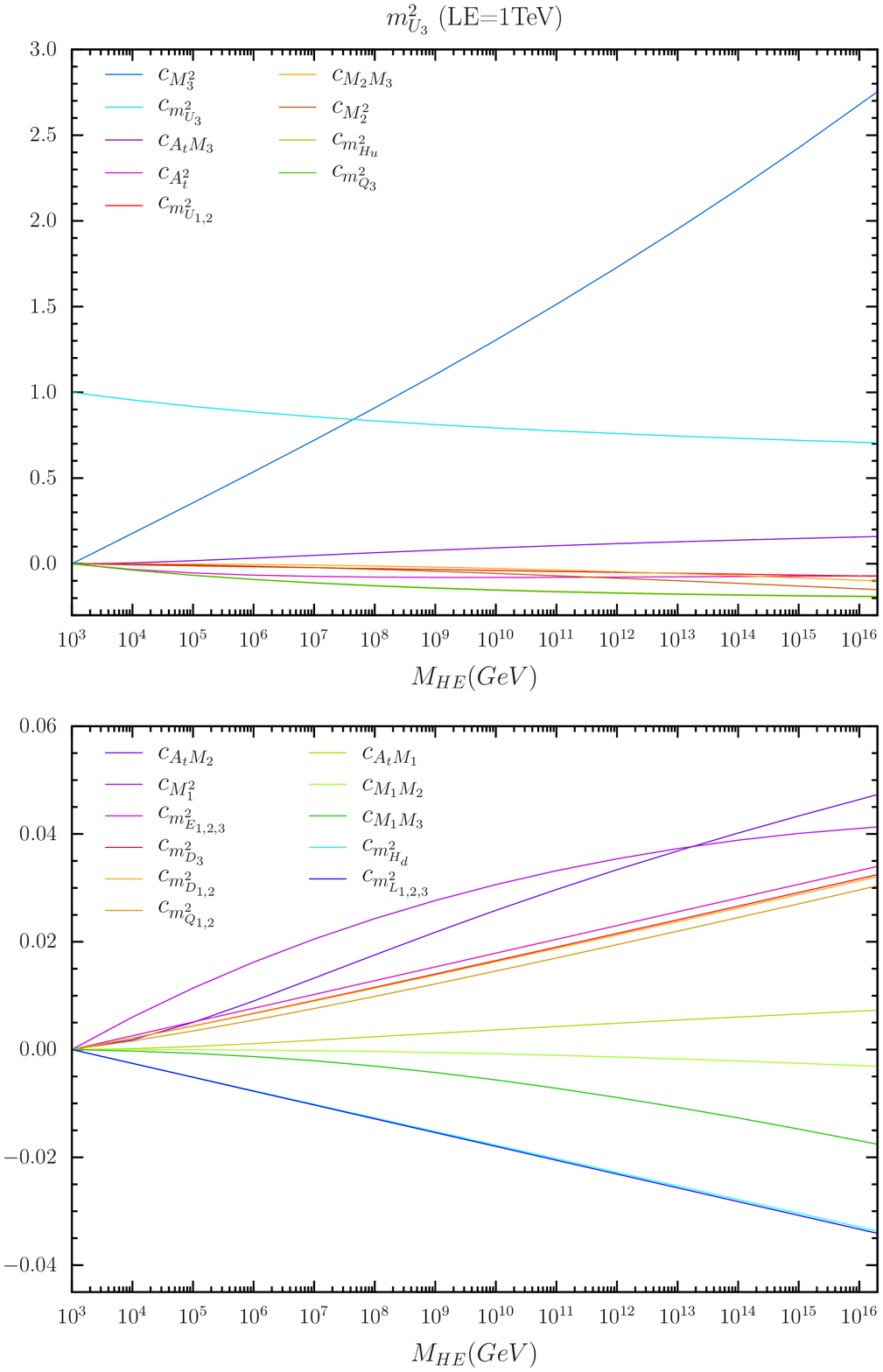}
\caption{As Figure 3, for $m_{U_3}^2(M_{\rm LE})$.} 
\label{fig:mU3sq}
\end{figure}

\begin{figure}[h!]
\centering 
\includegraphics[width=0.47\linewidth]{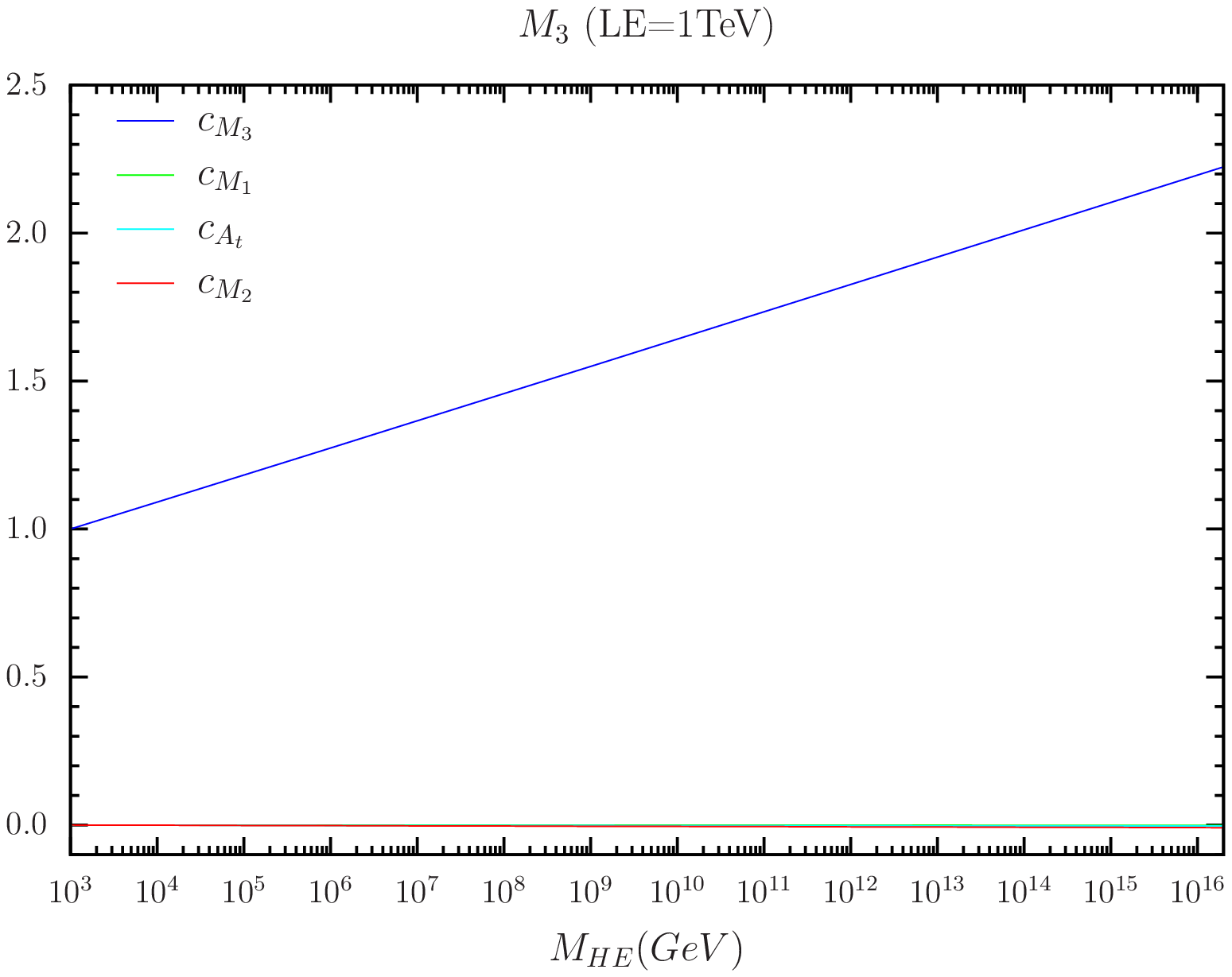}\hspace{0.5cm}
\includegraphics[width=0.47\linewidth]{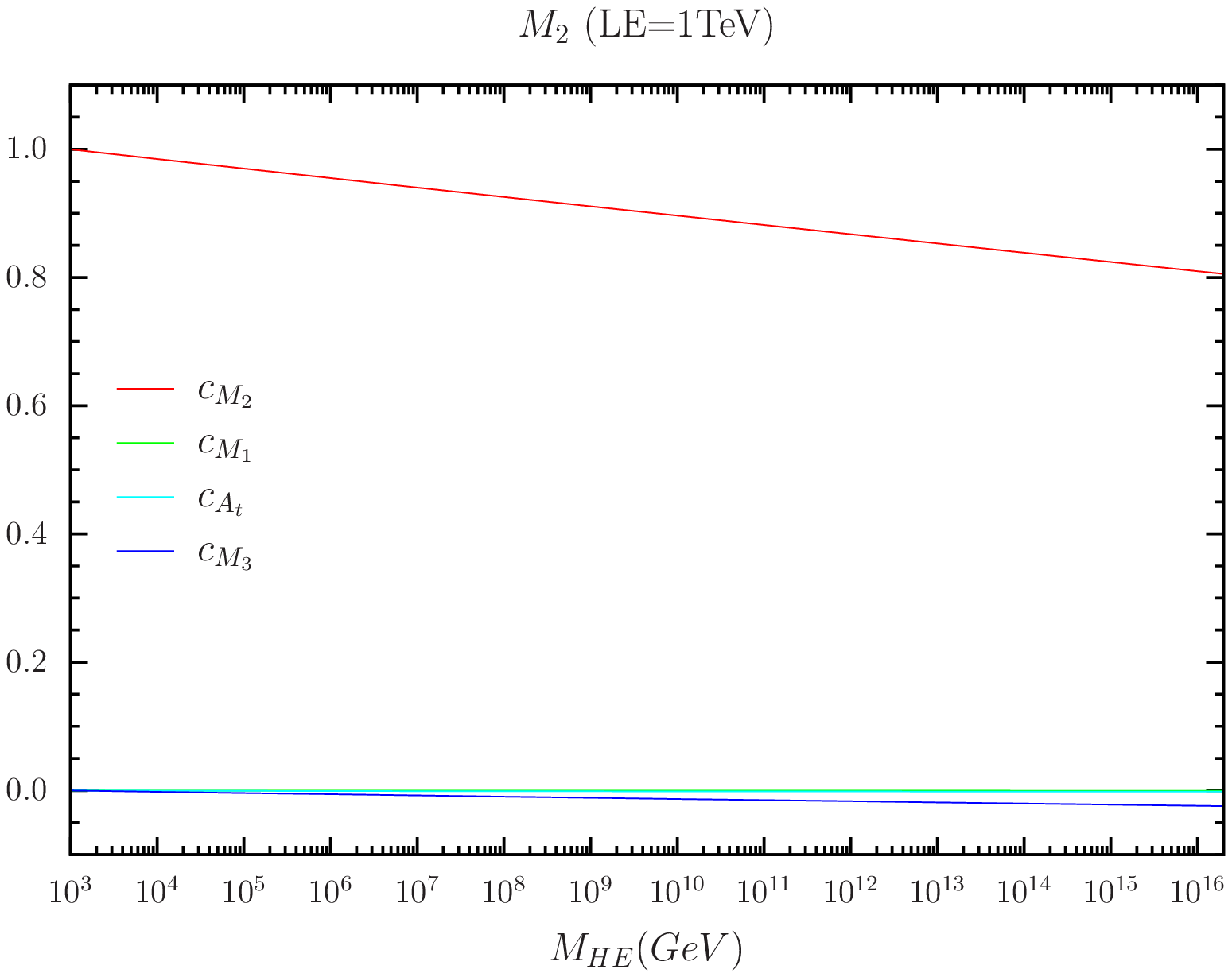}\\
\vspace{0.5cm} 
\includegraphics[width=0.47\linewidth]{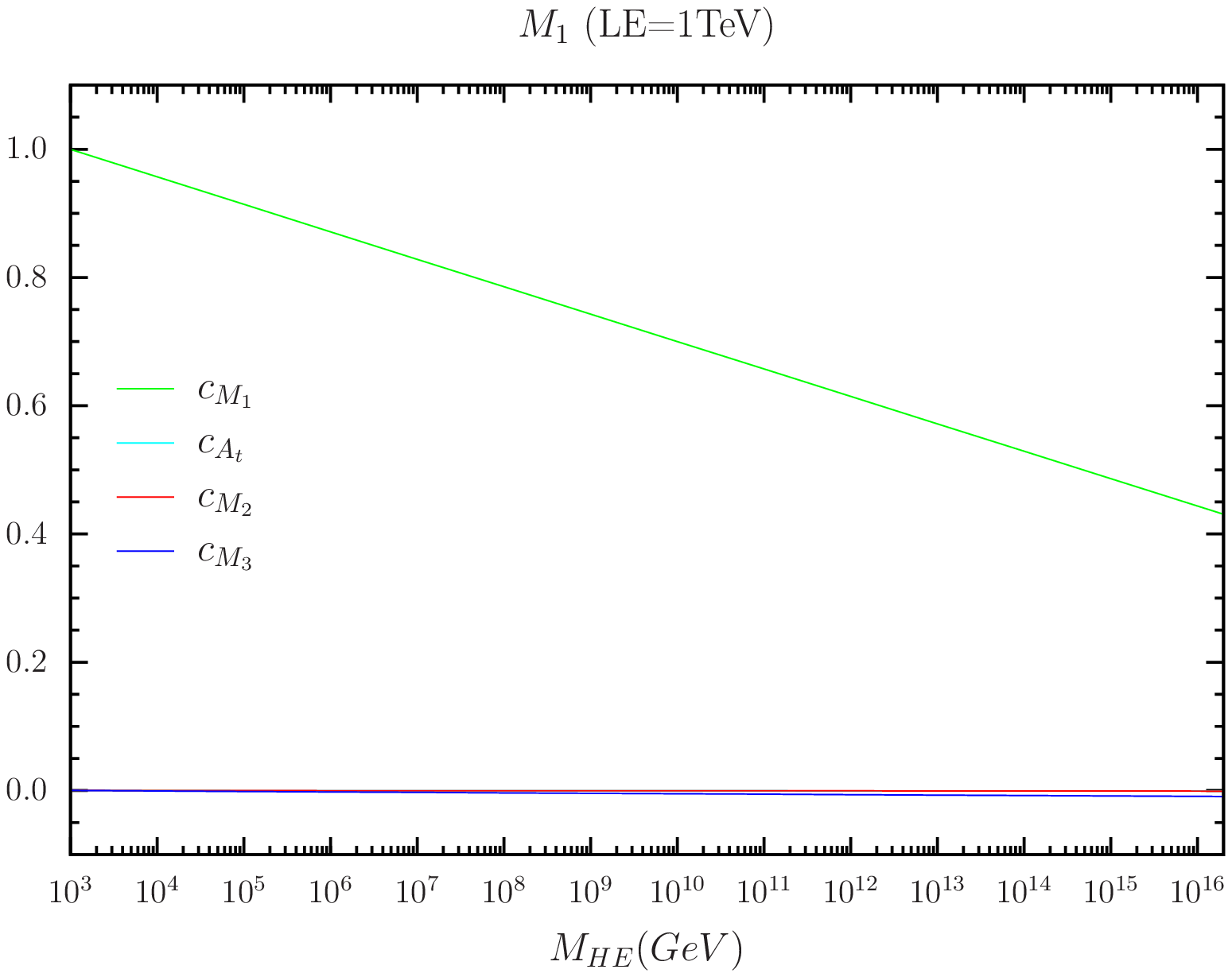}\hspace{0.5cm}
\includegraphics[width=0.47\linewidth]{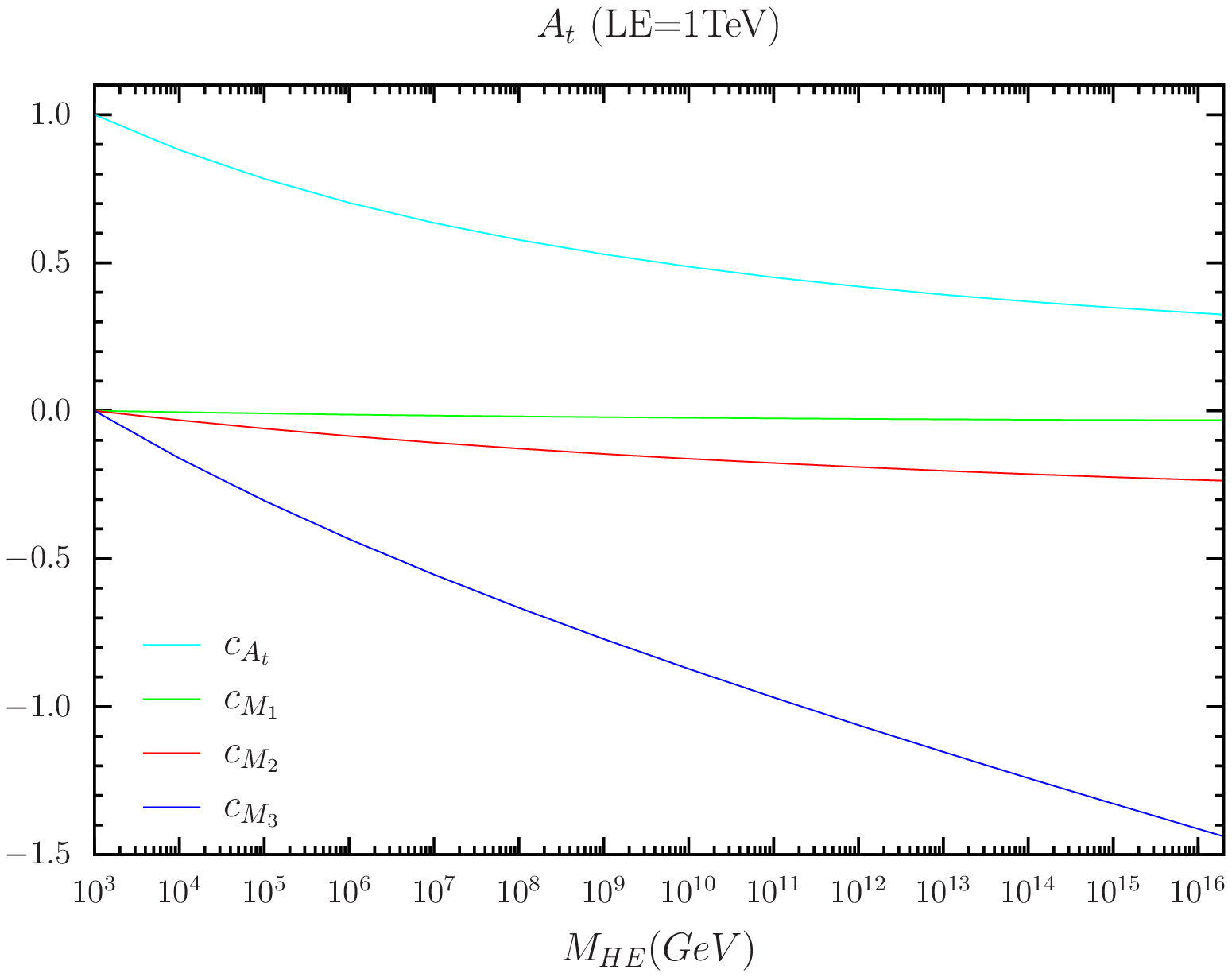}
\caption{Left to right, top to bottom: As Figure 3, for $M_3(M_{\rm LE})$, $M_2(M_{\rm LE})$, $M_1(M_{\rm LE})$ and $A_t(M_{\rm LE})$.} 
\label{fig:Mgauginos}
\end{figure}

\begin{figure}[h!]
\centering 
\includegraphics[width=0.6\linewidth]{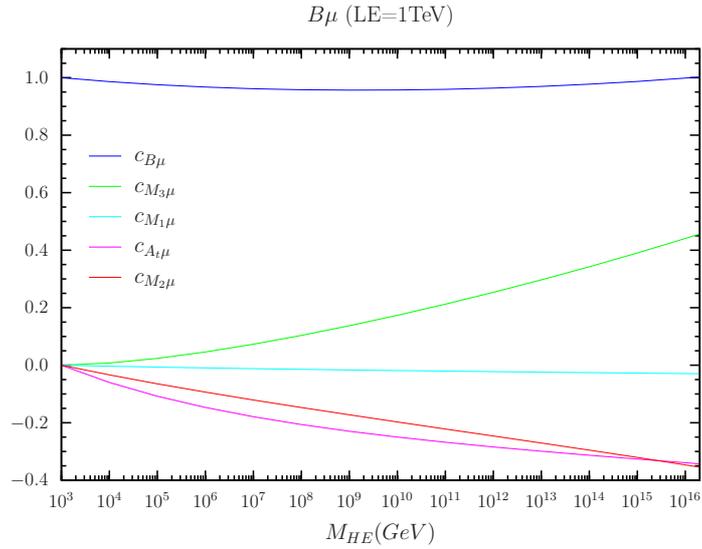}
\caption{As Figure 3, for $B\mu$.} 
\label{fig:Bmu}
\end{figure}

Finally, the dependence of the $c-$coefficients (and $b_{M_3^2}$) on $M_{\rm HE}$ is shown in Figs. \ref{fig:mHusq}, \ref{fig:mQ3sq}, \ref{fig:mU3sq}, \ref{fig:Mgauginos} and \ref{fig:Bmu}.

\begin{table}[h!]
 \centering
 {\small
\begin{tabular}[t]{|@{}l @{}l | S[table-format=2.4] | S[table-format=2.3] | S[table-format=2.4] | S[table-format=2.3] |}
\hline
{ } & {} & \multicolumn{2}{c|}{$m_{H_u}^2(M_{\rm LE})$} &  \multicolumn{2}{c|}{$m_{H_d}^2(M_{\rm LE})$}  \\
\cline{3-6}
{ } & {$HE$} & {$c_i$} & {$b_i$} & {$c_i$} & {$b_i$} \\
\hline
 & $M_3^2$ & -1.603 & 0.381 & -0.056 & 0.016 \\
 & $m_{H_u}^2$ & 0.631 & 0.019 & 0.025 & -0.001 \\
 & $m_{Q_3}^2$ & -0.367 & 0.018 & 0.015 &  {--} \\
 & $m_{U_3}^2$ & -0.290 & 0.017 & -0.051 & 0.001 \\
 & $A_tM_3$ & 0.285 & -0.024 & -0.002 & 0.001 \\
 & $M_2^2$ & 0.203 & 0.006 & 0.410 & -0.016 \\
 & $M_2M_3$ & -0.134 & 0.021 & -0.016 & 0.003 \\
 & $A_t^2$ & -0.109 & -0.006 & {--} & {--} \\
 & $A_tM_2$ & 0.068 &  {--} & -0.002 &  {--} \\
 & $m_{U_{1,2}}^2$ & 0.054 & -0.001 & -0.052 & 0.001 \\
 & $m_{H_d}^2$ & 0.026 & -0.001 & 0.961 & 0.001 \\
 & $m_{E_{1,2}}^2$ & -0.026 & 0.001 & 0.025 & -0.001 \\
 & $m_{E_3}^2$ & -0.026 & 0.001 & 0.023 & -0.001 \\
 & $m_{L_{1,2}}^2$ & 0.025 & -0.001 & -0.027 & 0.001 \\
 & $m_{L_3}^2$ & 0.025 & -0.001 & -0.029 & 0.001 \\
 & $m_{Q_{1,2}}^2$ & -0.025 &  {--} & 0.024 &  {--} \\
 & $m_{D_{1,2}}^2$ & -0.025 &  {--} & 0.026 & -0.001 \\
 & $m_{D_3}^2$ & -0.024 &  {--} & 0.016 &  {--} \\
 & $M_1M_3$ & -0.020 & 0.002 & -0.001 &  {--} \\
 & $A_tM_1$ & 0.012 &  {--} & {--} & {--} \\
 & $M_1^2$ & 0.006 & 0.002 & 0.033 &  {--} \\
 & $M_1M_2$ & -0.005 &  {--} & -0.001 &  {--} \\
 & $A_bM_3$ & -0.002 &  {--} & 0.022 & -0.005 \\
 & $A_b^2$ & 0.001 &  {--} & -0.009 & 0.001 \\
 & $A_bM_2$ & {--} & {--} & 0.006 & -0.001 \\
 & $A_{\tau}^2$ & {--} & {--} & -0.003 &  {--} \\
 & $A_{\tau}M_2$ & {--} & {--} & 0.002 &  {--} \\
 & $A_bA_t$ & {--} & {--} & 0.001 &  {--} \\
 & $A_{\tau}M_1$ & {--} & {--}& 0.001 &  {--} \\ 
 \hline
\end{tabular}
} 
\caption{$c_i$ and $b_i$ coefficients for the Higgs boson squared soft masses derived for $ \tan \beta = 10$, where `--' stands for HE parameters with $c_i, b_i < 0.001$. $M_{\rm LE}$  is set at 1~TeV.  For further details see eqs.~(\ref{m2_gen_fit}--\ref{scaledep_app}).}
\label{tab:mhiggssq}
\end{table}

\begin{table}[h!]
 \centering
 {\small
\begin{tabular}[t]{|@{}l @{}l |S[table-format=2.4] | S[table-format=2.3]| S[table-format=2.4] | S[table-format=2.3]| S[table-format=2.4] | S[table-format=2.3]| }
\hline
{ } & {} & \multicolumn{2}{c|}{$m_{Q_3}^2(M_{\rm LE})$} &  \multicolumn{2}{c|}{$m_{U_3}^2(M_{\rm LE})$} & \multicolumn{2}{c|}{$m_{D_3}^2(M_{\rm LE})$} \\
\cline{3-8}
{ } & {$HE$} & {$c_i$} & {$b_i$} & {$c_i$} & {$b_i$} & {$c_i$} & {$b_i$} \\
\hline
 & $M_3^2$ & 3.191 & -0.563 & 2.754 & -0.462 & 3.678 & -0.672 \\
 & $m_{Q_3}^2$ & 0.871 & 0.007 & -0.192 & 0.013  & -0.029 & 0.002 \\
 & $M_2^2$ & 0.333 & -0.008 & -0.151 & 0.017 & -0.010 & 0.002  \\
 & $m_{H_u}^2$ & -0.118 & 0.006 & -0.189 & 0.011 & -0.015 &  {--} \\
 & $m_{U_3}^2$ & -0.095 & 0.005 & 0.706 & 0.011 & 0.032 &  {--} \\
 & $M_2M_3$ & -0.084 & 0.015 & -0.100 & 0.018 & -0.026 & 0.007 \\
 & $A_tM_3$ & 0.072 & -0.003 & 0.159 & -0.010 & -0.010 & 0.003 \\
 & $A_t^2$ & -0.034 & -0.002 & -0.070 & -0.004 & 0.001 &  {--} \\
 & $A_tM_2$ & 0.020 &  {--} & 0.047 &  {--} & -0.001 &  {--} \\
 & $m_{Q_{1,2}}^2$ & -0.017 & 0.001 & 0.030 &  {--} & -0.025 & 0.002 \\
 & $m_{D_3}^2$ & -0.015 & 0.001 & 0.032 &  {--} & 0.973 & 0.001 \\
 & $m_{U_{1,2}}^2$ & 0.014 &  {--} & -0.073 & 0.002 & 0.031 &  {--} \\
 & $m_{D_{1,2}}^2$ & -0.012 & 0.001 & 0.032 &  {--} & -0.021 & 0.001 \\
 & $M_1M_3$ & -0.009 & 0.001 & -0.018 & 0.002 & -0.004 & 0.001 \\
 & $m_{E_{1,2,3}}^2$ & -0.009 &  {--} & 0.034 & -0.001 & -0.017 &  {--} \\
 & $m_{L_{1,2,3}}^2$ & 0.008 &  {--} & -0.034 & 0.001 & 0.017 &  {--} \\
 & $A_bM_3$ & 0.006 & -0.001 & -0.001 &  {--} & 0.014 & -0.003 \\
 & $M_1^2$ & -0.006 & 0.001 & 0.041 & 0.001 & 0.014 &  {--} \\
 & $m_{H_d}^2$ & 0.005 &  {--} & -0.034 & 0.001 & 0.011 &  {--} \\
 & $A_tM_1$ & 0.004 &  {--} & 0.007 &  {--} & {--} & {--} \\
 & $A_b^2$ & -0.003 &  {--} & {--} & {--} & -0.006 & 0.001\\
 & $M_1M_2$ & -0.002 &  {--} & -0.003 &  {--} & {--} & {--} \\
 & $A_bM_2$ & 0.002 &  {--} & {--} & {--} & 0.004 & -0.001 \\
 & $A_bA_t$ & 0.001 &  {--} & {--} & {--} & 0.001 &  {--} \\
 \hline
\end{tabular}
}
\caption{As table 3, for the squared soft masses of the third family squarks.}
\label{tab:msquarkthsq}
\end{table}

\begin{table}[h!]
 \centering
 {\small
\begin{tabular}[t]{|@{}l @{}l |S[table-format=2.4] | S[table-format=2.3]| S[table-format=2.4] | S[table-format=2.3]| S[table-format=2.4] | S[table-format=2.3]| }
\hline
{ } & {} & \multicolumn{2}{c|}{$m_{Q_1}^2(M_{\rm LE})$} &  \multicolumn{2}{c|}{$m_{U_1}^2(M_{\rm LE})$} & \multicolumn{2}{c|}{$m_{D_1}^2(M_{\rm LE})$} \\
\cline{3-8}
{ } & {$HE$} & {$c_i$} & {$b_i$} & {$c_i$} & {$b_i$} & {$c_i$} & {$b_i$} \\
\hline
 & $M_3^2$ & 3.672 & -0.674  & 3.702 & -0.680 & 3.708 & -0.681 \\
 & $m_{Q_1}^2$ & 0.982 & 0.001 & 0.028 &  {--} & -0.025 & 0.002 \\
 & $M_2^2$ & 0.403 & -0.015 & -0.005 & 0.001 & -0.005 & 0.001 \\
 & $M_2M_3$ & -0.046 & 0.009 & -0.022 & 0.006 & -0.021 & 0.006\\
 & $m_{Q_2}^2$ & -0.018 & 0.001 & 0.028 &  {--} & -0.025 & 0.002 \\
 & $m_{Q_3}^2$ & -0.017 & 0.001 & 0.029 &  {--} & -0.023 & 0.001 \\
 & $m_{U_3}^2$ & 0.015 &  {--} & -0.072 & 0.002 & 0.032 &  {--} \\
 & $m_{U_1}^2$ & 0.014 &  {--} & 0.927 & 0.002 & 0.031 &  {--} \\
 & $m_{U_2}^2$ & 0.014 &  {--} & -0.073 & 0.002 & 0.031 &  {--} \\
 & $m_{D_2}^2$ & -0.012 & 0.001 & 0.031 &  {--} & -0.021 & 0.001 \\
 & $m_{D_1}^2$ & -0.012 & 0.001 & 0.031 &  {--} & 0.979 & 0.001 \\
 & $m_{D_3}^2$ & -0.012 & 0.001 & 0.031 &  {--} & -0.021 & 0.001 \\
 & $m_{E_{1,2,3}}^2$ & -0.009 &  {--} & 0.034 & -0.001 & -0.017 &  {--} \\
 & $A_tM_3$ & -0.008 & 0.002 & -0.008 & 0.002 & -0.008 & 0.002 \\
 & $m_{H_u}^2$ & -0.008 &  {--} & 0.035 & -0.001 & -0.016 &  {--}  \\
 & $m_{H_d}^2$ & 0.008 &  {--} & -0.034 & 0.001 & 0.017 &  {--} \\
 & $m_{L_{1,2,3}}^2$ & 0.008 &  {--} & -0.034 & 0.001 & 0.017 &  {--} \\
 & $M_1M_3$ & -0.003 & 0.001 & -0.006 & 0.001 & -0.004 & 0.001  \\
 & $M_1^2$ & 0.003 &  {--} & 0.059 & -0.001 & 0.014 &  {--} \\
 & $A_tM_2$ & -0.001 &  {--} & {--} & {--} & {--} & {--} \\
 & $A_t^2$ & 0.001 &  {--} & 0.001 &  {--} & {--} & {--}\\
 \hline
\end{tabular}
}
\caption{As table 3, for the squared soft masses of the first family squarks.
Second generation squarks are degenerated with the first family.}
\label{tab:msquarkstsq}
\end{table}

\begin{table}[h!]
 \centering
 {\small
\begin{tabular}[t]{|@{}l @{}l |S[table-format=2.4] | S[table-format=2.3]| S[table-format=2.4] | S[table-format=2.3]| S[table-format=2.4] | S[table-format=2.3]| S[table-format=2.4] | S[table-format=2.3]| }
\hline
{ } & {} & \multicolumn{2}{c|}{$m_{L_3}^2(M_{\rm LE})$} &  \multicolumn{2}{c|}{$m_{E_3}^2(M_{\rm LE})$}  & \multicolumn{2}{c|}{$m_{L_1}^2(M_{\rm LE})$} &  \multicolumn{2}{c|}{$m_{E_1}^2(M_{\rm LE})$} \\
\cline{3-10}
{ } & {$HE$} & {$c_i$} & {$b_i$} & {$c_i$} & {$b_i$} & {$c_i$} & {$b_i$} & {$c_i$} & {$b_i$} \\
\hline
 & $m_{L_3}^2$ & 0.971 & 0.001 & 0.045 & -0.001 & -0.027 & 0.001 & 0.051 & -0.001 \\
 & $M_2^2$ & 0.416 & -0.017 & -0.004 &  {--} & 0.418 & -0.018 & {--} & {--}\\
 & $m_{U_{1,2}}^2$ & -0.052 & 0.001 & 0.104 & -0.002 & -0.052 & 0.001 & 0.104 & -0.002 \\
 & $m_{U_3}^2$ & -0.051 & 0.001 & 0.103 & -0.002 & -0.051 & 0.001 & 0.103 & -0.002 \\
 & $M_1^2$ & 0.034 &  {--} & 0.136 & -0.002 & 0.034 &  {--} & 0.137 & -0.002 \\
 & $m_{H_d}^2$ & -0.029 & 0.001 & 0.045 & -0.001 & -0.026 & 0.001 & 0.051 & -0.001 \\
 & $m_{L_1}^2$ & -0.027 & 0.001 & 0.051 & -0.001 & 0.973 & 0.001 & 0.051 & -0.001\\
 & $m_{L_2}^2$ & -0.027 & 0.001 & 0.051 & -0.001 & -0.027 & 0.001 & 0.051 & -0.001 \\
 & $m_{D_{1,2,3}}^2$ & 0.026 & -0.001 & -0.052 & 0.001 & 0.026 & -0.001 & -0.052 & 0.001 \\
 & $m_{E_1}^2$ & 0.025 & -0.001 & -0.052 & 0.001 & 0.025 & -0.001 & 0.948 & 0.001 \\
 & $m_{E_2}^2$ & 0.025 & -0.001 & -0.052 & 0.001 & 0.025 & -0.001 & -0.052 & 0.001 \\ 
 & $m_{H_u}^2$ & 0.025 &  {--} & -0.051 & 0.001 & 0.025 &  {--} & -0.051 & 0.001 \\
 & $m_{Q_3}^2$ & 0.024 &  {--} & -0.052 & 0.001 & 0.024 &  {--} & -0.052 & 0.001 \\
 & $m_{Q_{1,2}}^2$ & 0.024 &  {--} & -0.053 & 0.001 & 0.024 &  {--} & -0.053 & 0.001 \\
 & $m_{E_3}^2$ & 0.023 & -0.001 & 0.942 & 0.001 & 0.025 & -0.001 & -0.052 & 0.001 \\
 & $M_2M_3$ & -0.009 & 0.001 & 0.001 &  {--} & -0.009 & 0.001 & 0.001 &  {--} \\
 & $M_3^2$ & -0.007 & 0.001 & -0.001 &  {--} & -0.007 & 0.001 & -0.001 &  {--} \\
 & $A_{\tau}^2$ & -0.003 &  {--} & -0.006 &  {--} & {--} & {--} & {--} & {--} \\
 & $A_{\tau}M_2$ & 0.002 &  {--} & 0.003 &  {--} & {--} & {--} & {--}& {--}\\
 & $A_tM_2$ & -0.001 &  {--} & {--} & {--} & -0.001 &  {--} & {--}& {--} \\
 & $M_1M_2$ & -0.001 &  {--} & -0.001 &  {--} & -0.001 &  {--} & {--} & {--} \\
 & $A_{\tau}M_1$ & 0.001 &  {--} & 0.001 &  {--} & {--} & {--} & {--} & {--} \\
 & $M_1M_3$ & {--} & {--} & -0.002 &  {--} & {--} & {--} & -0.002 &  {--} \\
 & $A_tM_3$ & {--} & {--} & -0.001 &  {--} & {--} & {--} & -0.001 &  {--} \\
 \hline
\end{tabular}
}
\caption{As table 3,  for the squared soft masses of the third and first family sleptons. Second generation sleptons are  degenerated with the first family.}
\label{tab:msqslept}
\end{table}

\begin{table}[h!]
 \centering
 {\small
\begin{tabular}[t]{|@{}l @{}l |S[table-format=2.4] | S[table-format=2.3]| S[table-format=2.4] | S[table-format=2.3]| S[table-format=2.4] | S[table-format=2.3]| }
\hline
{ } & {} & \multicolumn{2}{c|}{$M_3(M_{\rm LE})$} &  \multicolumn{2}{c|}{$M_2(M_{\rm LE})$} & \multicolumn{2}{c|}{$M_1(M_{\rm LE})$} \\
\cline{3-8}
{ } & {$HE$} & {$c_i$} & {$b_i$} & {$c_i$} & {$b_i$} & {$c_i$} & {$b_i$} \\
\hline
 & $M_3$ & 2.224 & -0.160 & -0.024 & 0.004 & -0.009 & 0.001 \\
 & $M_2$ & -0.009 & 0.001 & 0.806 & 0.011 & -0.001 &  {--} \\
 & $A_t$ & -0.003 &  {--} & -0.002 &  {--} & -0.001 &  {--} \\
 & $M_1$ & -0.001 &  {--} & {--} & {--} & 0.431 & 0.012 \\
\hline
\end{tabular}
}
\caption{As table 3,  for the gaugino masses. }
\label{tab:gauginos}
\end{table}

\begin{table}[h!]
 \centering
 {\small
\begin{tabular}[t]{|@{}l @{}l |S[table-format=2.4] | S[table-format=2.3]| S[table-format=2.4] | S[table-format=2.3]| S[table-format=2.4] | S[table-format=2.3]| }
\hline
{ } & {} & \multicolumn{2}{c|}{$A_t(M_{\rm LE})$} &  \multicolumn{2}{c|}{$A_b(M_{\rm LE})$} & \multicolumn{2}{c|}{$A_\tau(M_{\rm LE})$} \\
\cline{3-8}
{ } & {$HE$} & {$c_i$} & {$b_i$} & {$c_i$} & {$b_i$} & {$c_i$} & {$b_i$} \\
\hline
 & $M_3$ & -1.438 & 0.148 & -2.129 & 0.277 & 0.017 & -0.003 \\
 & $A_t$ & 0.325 & 0.035 & -0.106 & 0.005 & 0.001 &  {--} \\
 & $M_2$ & -0.237 & -0.005 & -0.413 & 0.016 & -0.460 & 0.022 \\
 & $M_1$ & -0.032 & -0.002 & -0.030 & {--} & -0.145 & 0.003\\
 & $A_b$ & -0.002 &  {--} & 0.981 & 0.002 & -0.010 &  0.001 \\
 & $A_{\tau}$ & {--} & {--} & -0.003 &  {--} & 0.988 & {--}  \\
\hline
\end{tabular}
}
\caption{As table 3, 
for the trilinear scalar couplings.}
\label{tab:trilinears}
\end{table}

\begin{table}[h!]
 \centering
\begin{minipage}[t]{.4\linewidth}
 \centering
 {\small
\begin{tabular}[t]{|@{}l @{}l |S[table-format=2.4] | S[table-format=2.3]| }
\hline
{ } & {} & \multicolumn{2}{c|}{$\mu(M_{\rm LE})$} \\
\cline{3-4}
{ } & {$HE$} & {$c_i$} & {$b_i$} \\
\hline
 & $\mu$ & 1.002 & 0.013 \\
\hline
\end{tabular}
}
\end{minipage}%
\begin{minipage}[t]{.4\linewidth}
 \centering
 {\small
\begin{tabular}[t]{|@{}l @{}l |S[table-format=2.4] |  S[table-format=2.3]| }
\hline
{ } & {} & \multicolumn{2}{c|}{$B\mu(M_{\rm LE})$} \\
\cline{3-4}
{ } & {$HE$} & {$c_i$} & {$b_i$} \\
\hline
 & $B\mu $ & 1.002 & 0.013 \\
 & $M_3\mu$  & 0.456 & -0.080 \\
 & $M_2\mu$  & -0.354 & 0.004 \\
 & $A_t\mu$  & -0.343 & 0.013 \\
 & $M_1\mu$  & -0.030 & -0.001 \\
 & $A_b\mu$  & -0.009 & 0.001 \\
 & $A_{\tau}\mu$  & -0.003 &  {--} \\
\hline
\end{tabular}
}
\end{minipage} 
\caption{Left, $c_i$ and $b_i$ coefficients for the $\mu-$parameter. Right, $c_i$ and $b_i$ coefficients for $B\mu$.}
\label{tab:mu-Bmu}
\end{table}

\newpage

\FloatBarrier

\bibliographystyle{JHEP}    
\bibliography{CMRRZ}	 

\end{document}